\documentclass[letterpaper]{ar-1col} 
\usepackage[numbers]{natbib}
\pdfoutput=1

\jname{Annu. Rev. Astron. Astrophys.}
\jvol{54}
\jyear{2016}
\doi{10.1146/((please add article doi))}

\newcommand{\smy}{\,M$_\odot$\,yr$^{-1}$}
\newcommand{\ha}{H$\alpha$}
\newcommand{\kms}{\,km/s} 
\newcommand{\msun}{\,M$_\odot$}
\newcommand{\sqcm}{\,cm$^{-2}$}

\begin{document}

\markboth{D'Onghia and Fox}{The Magellanic Stream}

\title{The Magellanic Stream: Circumnavigating the Galaxy}

\author{Elena D'Onghia,$^1$ and Andrew J. Fox,$^2$
\affil{$^1$Astronomy Department, University of Wisconsin, 475 N Charter
  Street, Madison, WI, 53706; edonghia@astro.wisc.edu}
\affil{$^2$Space Telescope Science Institute, 3700 San Martin Drive, 
Baltimore, MD, 21218; afox@stsci.edu}}

\begin{abstract}
The Magellanic Clouds are surrounded by an extended network of gaseous 
structures. Chief among these is the Magellanic Stream, an interwoven tail 
of filaments trailing the Clouds in their orbit around the Milky Way.
When considered in tandem with its Leading Arm, the Stream stretches 
over 200 degrees on the sky. Thought to represent the result of tidal 
interactions between the Clouds and ram-pressure forces exerted by the 
Galactic corona, its kinematic properties reflect
the dynamical history of the closest pair of dwarf galaxies 
to the Milky Way. The Stream is a benchmark for hydrodynamical simulations 
of accreting gas and cloud/corona interactions. If the Stream survives 
these interactions and arrives safely in the Galactic disk, its cargo of 
over a billion solar masses of gas has the potential to maintain or 
elevate the Galactic star formation rate. In this article, we review
the current state of knowledge of the Stream, including its chemical 
composition, physical conditions, origin, and fate.  
We also review the dynamics of the Magellanic System, including
the proper motions and orbital history of the Large and Small Magellanic
Clouds, the first-passage and second-passage scenarios,
and the evidence for a Magellanic Group of galaxies. 
\end{abstract}

\begin{keywords}
Magellanic Clouds; galaxy interactions, tidal stripping, circumgalactic gas,
gas kinematics, accretion, dynamics  
\end{keywords}
\maketitle

\tableofcontents

\section{INTRODUCTION}

\begin{itemize}
\item [] {\it ``If we worked on the assumption that what is accepted as true really is true, then there would be little hope for advance'' -- Orville Wright, aviation pioneer} 
\end{itemize}

The Magellanic Stream, an extended tail of neutral and ionized gas
trailing the Magellanic Clouds in their orbit around the Milky Way (MW),
was discovered in several radio surveys of the southern sky.
The first hints of its existence were provided by 
Dieter (1965), who mapped high Galactic latitude regions
in 21 cm emission and found a population of high-velocity
clouds (HVCs) near the south Galactic pole. 
Other H I clouds were soon found in similar regions of the sky
(Hulsbosch \& Raimond 1966, Hulsbosch 1968, Dieter 1971), and while 
the distance to these clouds was 
at that time unknown, the deeper 21 cm studies of 
van Kuilenburg (1972) and Wannier \& Wrixon (1972)
revealed they formed a coherent linear structure.
Soon afterwards Mathewson et al. (1974) used the Parkes
radio telescope to trace the structure back to the Magellanic Clouds, 
and coined the term Magellanic Stream (hereafter the Stream).
The Mathewson et al. paper is often credited with the discovery of the Stream; 
in hindsight, all the H I papers cited above had detected parts of it, 
but Mathewson et al. were the first to demonstrate its Magellanic source.
Mathewson et al. 1974 further detected a counterpart to the Stream
on the other side of the Magellanic Clouds (the Leading Arm)
that extends across the Galactic plane into positive Galactic latitudes, and 
suggested the extension was a continuation of the same structure.
The Magellanic nature of the Leading Arm was later
confirmed by kinematic (Putman et al. 1998) and
metallicity (Lu et al. 1998) measurements.

Since this initial 
phase of discovery and characterization, 
other methods of studying the Stream
have become available, notably ultraviolet (UV) and optical 
absorption-line spectroscopy of 
background sources, \ha\ emission-line spectroscopy, $N$-body modeling,
and hydrodynamical simulations. 
The UV and optical spectroscopic studies are complementary to the radio studies
since they address the ionized gas in the Stream. At the same time,
successively more sensitive radio surveys at 21 cm have been completed,
bringing the filamentary and fragmented structure of the Stream into focus.

Together these approaches have revolutionized our understanding 
of the Stream. 
We now know it is a bifurcated, multi-phase, fragmented, 
massive gaseous structure that originated in both Magellanic Clouds.
It has become a benchmark for the study of gas-transport processes
in the halo of an $L_*$-galaxy, and
is the only known {\it gaseous} tidal stream in the vicinity of the 
MW, where {\it stellar} tidal streams are more common. 
It dominates all the other gaseous high-velocity clouds (HVCs) in the 
Galactic halo, both in terms of gas mass and inflow rate,
and therefore understanding the Stream is essential to a
global picture of the Galaxy's circumgalactic medium.
Its age and location on the sky constrain dynamical models of the
Magellanic Clouds and the total mass of the MW.
These many independent reasons have turned the attention of various groups 
of astronomers onto the Stream. In this review article we synthesize current 
knowledge and lay out a path for future studies. We also review
the subject of {\it Magellanic Dynamics}, which is closely related
since the Stream's location traces out the past orbits of the 
LMC and SMC.

We use the term \emph{The Magellanic System} to include the LMC, SMC, and 
the three notable gaseous structures surrounding them: the Stream, Bridge, 
and Leading Arm.
The masses of the various components of the
Magellanic System are given in \textbf {Table \ref{msys}}.
Note that all gas masses are given assuming a distance $d$=55 kpc,
but will scale as $d^2$.

\begin{table}    
\caption{The Magellanic System: Mass Budget}
\label{msys}    
\begin{tabular}{lll lllr} 
\hline
Property & LMC & SMC & Stream & Bridge & Leading Arm & Refs.\\
\hline
Stellar mass          & 3$\cdot$10$^9$\msun\    & 3$\cdot$10$^8$\msun\ 
& ... & 1.5$\cdot$10$^4$\msun$^a$ & ... & (1,2,3) \\
Halo mass     & 1.7$\cdot$10$^{10}$\msun$^b$     & 2.4$\cdot$10$^{9}$\msun$^c$ 
& ...& ... & ... & (4,2)\\
H I gas mass          & 4.4$\cdot$10$^8$\msun\    & 4.0$\cdot$10$^8$\msun\ 
& 2.7$\cdot$10$^8$\msun$^d$  & 1.8$\cdot$10$^8$\msun\  & 3.0$\cdot$10$^7$\msun\ 
& (5)\\
H II gas mass         & ... & ... & 
$\approx$10$^9$\msun$^e$ & 0.7--1.7$\cdot$10$^8$\msun$^f$ & $\sim$10$^8$\msun$^e$
& (6,7)\\
\hline
\end{tabular}   
\begin{tabnote}
References: (1) van der Marel et al. 2002; (2) Stanimorovi\'c et al. 2004;
(3) Harris 2007, (4) van der Marel \& Kallivayalil 2014; 
(5) Br\"uns et al. 2005; (6) Fox et al. 2014; (7) Barger et al. 2013.\\
$^a$ Assuming a Kroupa IMF and a 10\,Gyr old stellar population.\\
$^b$ Out to a radius 8.7\,kpc; total value depends on tidal truncation 
of LMC halo.\\
$^c$ Dynamical mass based on H I rotation curve. See also Bekki \& 
Stanimirovi\'c 2009.\\
$^d$ Mass summed over MS regions I--IV and Interface Region, which we treat 
as part of Stream. Assumes a distance $d$=55 kpc.\\
$^e$ Based on UV absorption. Includes warm-ionized ($\sim$10$^4$ K)
and hot-ionized ($\sim$10$^5$ K) contributions.\\
$^f$ Based on \ha\ emission. Includes warm-ionized contribution.
\end{tabnote}
\end{table}

\begin{marginnote}
\entry{LA}{Leading Arm}
\entry{MB}{Magellanic Bridge}
\entry{MS}{Magellanic Stream}
\entry{MW}{Milky Way}
\end{marginnote}

\section{THE MAGELLANIC SYSTEM: OBSERVED PROPERTIES}
\label{ms-obsprop}

In this section the basic observational properties of the Magellanic
System are described, 
starting with an overview of the Magellanic Clouds (Sect. \ref{overview}).
We then cover the Magellanic Stream (Sect. \ref{streamsection}),
including its structure and morphology, chemical abundance patterns, 
dust properties, and ionized gas content. 
We then present an overview of the 
Magellanic Bridge (Sect. \ref{bridge}) and Leading Arm (Sect. \ref{la}).
Finally we discuss the total mass and mass-flow rate of Magellanic gas 
in Sect. \ref{totalmass}. 

\subsection{The Magellanic Clouds}
\label{overview}

The Large and Small Magellanic Clouds (LMC/SMC)
are dwarf irregular galaxies 
in the southern hemisphere. Both are satellite galaxies of the MW, 
visible to the naked eye 
in the constellations Dorado (LMC) and Tucana (SMC). 
They have therefore been known since ancient times, but adopt their name 
from Ferdinand Magellan who observed them on his expedition in 1519--1522. 
The Clouds are the closest pair of galaxies to the MW
and a unique location for studying
galaxy-formation processes. The Clouds and the MW form the nearest 
ensemble of interacting galaxies, representing 
the best opportunity anywhere for
studying satellite-satellite and satellite-host galaxy
gravitational interactions.

The LMC, at a distance of 50$\pm$1 kpc 
(Pietrzy{\'n}ski et al. 2013, Walker 2012),
is the prototypical Barred Magellanic Spiral, a population
of galaxies with peculiar morphological properties. These include 
an asymmetric stellar bar, which may be off-centered from the 
dynamical center of the galaxy, a single looping spiral arm, 
and often a large star-forming complex at one end of the bar 
(de Vaucouleurs \& Freeman 1972). 
The LMC's stars are concentrated into a flat disk tilted
at an inclination $i$=45$^o$ with respect to the plane of the sky, and it
has a luminosity one-tenth that of the MW
(Sparke \& Gallagher 2000).
Its current-day metallicity, as determined from massive main-sequence 
stars, is [Z/H]=$-$0.31$\pm$0.04, i.e. $\approx$0.5 solar 
(Rolleston et al. 2002; see also Russell \& Dopita 1992).
The total LMC mass (out to 8.7 kpc) is 1.7$\cdot$10$^{10}$\msun\
(van der Marel \& Kallivayalil 2014; see Table 1), although it is unclear
how extended the halo of LMC is, so the total mass 
could be considerably higher.

The SMC, at 61$\pm$1 kpc (Hilditch et al. 2005, Graczyk et al. 2014), 
is the closest dwarf irregular galaxy, and an excellent benchmark for 
studying the evolution of late-type dwarf galaxies. 
It has an elongated (cigar-shaped) structure seen end-on.
The SMC's current-day metallicity is
$\approx$20\% solar derived from H II regions and young stars 
(Russell \& Dopita 1992), 
although the red giant population
has [Fe/H]=$-$0.99$\pm$0.01, i.e. 10\% solar (Dobbie et al. 2014).
The total SMC mass is estimated from rotation curves
to be 2.4$\cdot$10$^9$\msun\ (Stanimirovi\'c et al. 2004),
at the upper limit of the mass range of this
class of galaxies (Tolstoy et al. 2009). 
Due to its strong gravitational interaction with the LMC,
the SMC exhibits a complex morphology and dynamics,
with the gas behaving differently from the stellar populations.
The large-scale H I distribution
shows an irregular and asymmetric morphology, with
prominent features including the visual appearance of a bar, possible
extensions to the bar similar to arms, 
and a ``wing'' connected by a bridge of gas
(Caldwell \& Coulson 1986; Stanimirovi\'c et al. 2004). 
All of these features might be signs of recent tidal
interactions with the LMC (Maragoudaki et al. 2001). The small-scale H I
distribution is characterized by numerous arcs, filaments,
and expanding shells (Stanimirovi\'c et al. 1999). 
The old and young stellar populations in the SMC
have different spatial distributions (e.g. Gardiner \& Hatzidimitriou 1992). 
Young stars with ages $\le$200 Myr have a distribution that follows
the H I gas (Zaritsky et al. 2000), whereas the old stars with 
ages $\ge$1 Gyr show a regular spheroidal distribution 
(Cioni et al. 2000, Harris \& Zaritsky 2006).

The fact that the Magellanic Clouds are so well studied and characterized
provides clear advantages for understanding the Stream, since we can compare
the kinematics, chemistry, and dust content of the Stream with the
corresponding properties of its parent galaxies, and thereby refine our 
understanding of the Stream's origin.

\subsection{The Magellanic Stream}
\label{streamsection}

\subsubsection{The neutral gas: structure and morphology}
\label{morphology}

\begin{figure}
\includegraphics[width=5.0in]{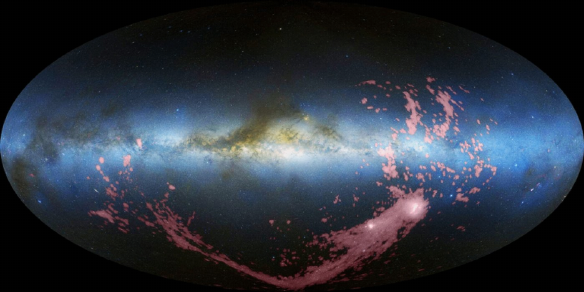}  
\caption{The Magellanic Stream (pink) displayed in Galactic coordinates using
an all-sky Hammer-Aitoff projection centered on the Galactic Center 
(image credit Nidever et al. 2010, NRAO/AUI/NSF, Mellinger 2009, 
Leiden-Argentine-Bonn Survey, Parkes Observatory, Westerbork Observatory, 
Arecibo Observatory).
The Magellanic Bridge between the LMC and SMC and the Leading Arm are 
clearly visible, as is the extended fragmentation across the Stream.}
\label{fig1}
\end{figure}

The Stream was discovered and characterized in H I 21 cm emission, 
and so we have a detailed understanding of the
spatial structure and morphology of the \emph{neutral} gas.
Our knowledge has progressed hand-in-hand with
successive generations of radio telescopes. Major 21 cm surveys
covering parts or all of the Stream include those by 
Mathewson et al. (1974, 1977), Bajaja et al. (1985),
Hulsbosch \& Wakker (1988), Morras et al. (2000), Lockman et al. (2002),
Putman et al (2003b), Br\"uns et al. (2005), Kalberla et al. (2005),
Stanimirovi\'c et al. (2008),
Nidever et al. (2008, 2010), and McClure-Griffiths et al. (2009)
The full size of the Stream can be seen in \textbf{Figure \ref{fig1}} 
(Nidever et al. 2010), which shows its extent on an all-sky map.

\begin{figure}
\includegraphics[width=5.0in]{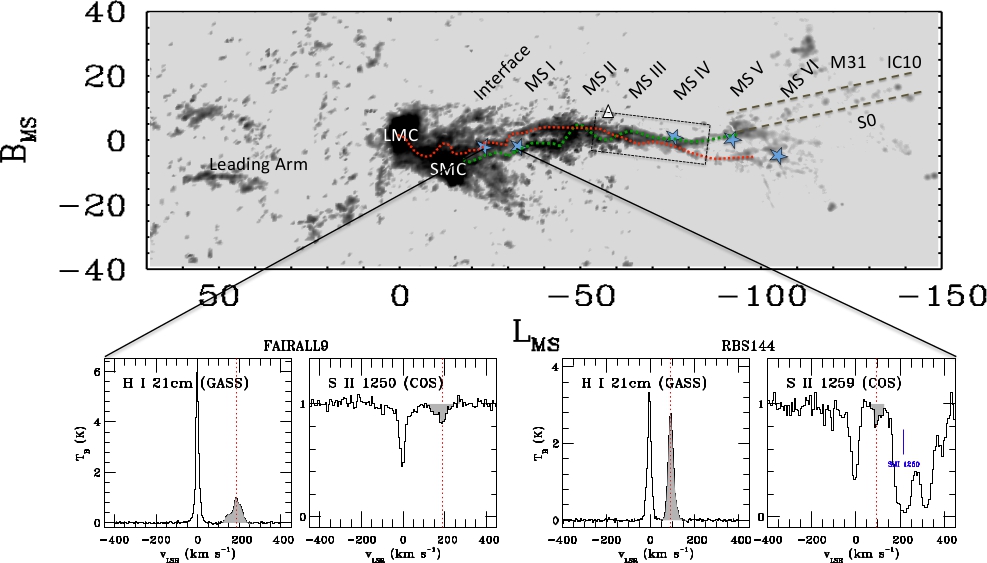} 
\caption{
{\bf Top:} Overview of the Magellanic System (Hammer et al. 2015), 
showing H I data from the GASS survey (McClure-Griffiths et al. 2009)
plotted in the Magellanic coordinate system of Nidever et al. (2010).
The main components of the system are labeled. The red and green
dotted lines denote the LMC and SMC filaments, respectively, and the blue
stars show directions where UV metal-abundance measurements are available
(Fox et al. 2013).
{\bf Bottom:} 21 cm and UV spectra of the Stream in two directions,
one passing through the SMC filament (RBS 144) and one passing through
the LMC filament (Fairall 9). The relative strength
of the H I emission and S II absorption provides a direct measurement
of the metallicity. The two filaments have sulfur abundances differing
by a factor of five, confirming their different origin 
(Fox et al. 2013; Richter et al. 2013).}
\label{overviewfig}
\end{figure}

Mathewson et al. (1977) identified six horseshoe-shaped concentrations of 
gas along the Stream, which they named MS I--VI with increasing 
angular separation from the Clouds. The gas surface density decreases 
monotonically from MS I and to MS VI. This naming convention 
is still seen in the literature, but the higher sensitivity of the more 
recent surveys has revealed diffuse gas between the dense clumps.
We now know that the Stream consists of an almost linear 
central body surrounded by a significant number of small-scale fragmentary 
structures. The small-scale structure is evident across the Stream, from 
its tip farthest from the Magellanic Clouds to its source where it 
connects with them (Mirabel et al. 1979, Haynes 1979, Mirabel 1981, 
Wayte 1989, Wakker et al. 2002, Stanimirovi\'c et al. 2002, 2008, 
Putman et al. 2003b, Westmeier \& Koribalski 2008). 
The Leading Arm (LA) also shows substantial small-scale structure 
(Putman et al. 1998).
The extent of the fragmentation is revealed by the work of For et al. (2014),
who cataloged 251 discrete clouds and filaments across the Stream.
In {\bf Figure \ref{overviewfig}} we show an overview of the H I emission from
the Stream, with the various components labeled.

Furthermore, the central body of the Stream is \emph{bifurcated} into two 
main strands or filaments (see Figure \ref{overviewfig}), 
as was first recognized by Cohen (1982) and Morras (1983).
Deeper observations of the Stream's bifurcated structure were presented by
Putman (2000) and Putman et al. (2003b), using data from the H I Parkes 
All Sky Survey (HIPASS; Barnes et al. 2001). 
The bifurcation of the Stream has also been detected {\it kinematically} 
(Nidever et al. 2008), in that the two strands have velocities that trace 
back to different origins (with one connecting to the LMC), and 
{\it chemically} (Fox et al. 2013, Richter et al. 2013), in that the two
strands show different chemical abundance patterns, one resembling the SMC
pattern and the other resembling the LMC pattern.
The Stream is therefore an interwoven tail of two filaments.

The kinematic bifurcation found by
Nidever et al. (2008) was derived from a Gaussian decomposition of the 
Leiden-Argentine-Bonn (LAB) 21\,cm emission data on the Stream
(Kalberla et al. 2005). Using position-velocity diagrams, Nidever et al.
showed that one filament of the Stream can be traced back 
to the LMC, not the SMC, as had traditionally been assumed. Furthermore, 
this LMC filament connects to a region of the LMC known as the southeast 
H I over-density (SEHO), which includes the starburst region 30 Doradus. 
This provides an important and hitherto-unknown connection
between the Stream and star-formation activity within the Clouds,
as if the LMC strand is a smoke trail from a Magellanic chimney.
Nidever et al. report that the other filament is not conclusively connected
to either Magellanic Cloud, but its chemical properties have subsequently
been shown to indicate an SMC origin (see Sect. \ref{abundances}).

Many bulk properties of the neutral gas in the Stream are well known.
The Stream has a total H I mass of 2.7$\cdot$10$^{8}$\msun\
(for a distance $d$=55\,kpc, and including the ``Interface Region''
near the SMC; Br\"uns et al. 2005).
When considered in tandem with the Bridge and Leading Arm, 
the total H I mass in the Magellanic System is 5.0$\cdot$10$^{8}$\msun\
and covers a total angular extent of 2700 square degrees on the sky 
down to a limiting column density $N$(H I)=10$^{18.0}$\sqcm\ 
(Nidever et al. 2010).
There is a fairly smooth column density gradient along the Stream
(Mathewson et al. 1974, Putman et al. 2003b, Nidever et al. 2010)
which can be written as
$N$(H I)=5.9$\cdot$10$^{21}$exp($l_{\rm MS}$/19.3$^\circ$)\sqcm,
where $l_{\rm MS}$ is the Magellanic Longitude, the angular distance
from the LMC along the main axis of the Stream 
(Nidever et al 2010; see also Wakker 2001).
The velocity of the Stream in the Local Standard of Rest (LSR)
varies from +180 km/s near the Clouds to $-$450\kms\ near its tip.

Nidever et al. (2008) reported periodic, sinusoidal 
undulations in the H I Stream; they suggest
these could be the imprint of the LMC rotation curve, related to the position
of the SEHO in the LMC at the time the gas was stripped.
In this scenario the drift rate of the Stream away from the LMC
is $\sim$49\kms\ and its age is $\sim$1.7 Gyr, in good agreement
with tidal age estimates of the Stream (Besla et al. 2010, 2012). 
With the more spatially-extended Stream reported by 
the deeper H I measurements of Nidever et al. (2010), the age increases to 
$\sim$2.5 Gyr. Putman et al. (2003b) had already noticed the 
double-helix morphology of the Stream and suggested it may be related
to the binary motion of the LMC and SMC.

The typical H I line widths in the Stream are 18--25\kms\ 
(FWHM; Mathewson \& Ford 1984; Hsu et al. 2011), 
which under the assumption of thermal broadening correspond to 
gas temperatures of 7000--14000\,K, although there
are regions with line widths as narrow as 3 km/s
(Kalberla \& Haud 2006, Stanimirovi\'c et al. 2008).
Regions of cooler gas in the Stream were 
detected by Matthews et al. (2009), via 21\,cm
absorption studies. They detected a cool cloud with an inferred
spin temperature of 70\,K toward one continuum source, but failed to detect
cool gas toward three other sources, indicating the cool phase
has a low covering fraction. 
Furthermore, a UV detection of H$_2$ absorption by Richter et al. (2001)
toward the QSO Fairall 9
with a total H$_2$ column log\,$N$(H$_2$)=16.40$^{+0.28}_{-0.53}$ and a molecular
fraction $f$(H$_2$)=5.4$\cdot$10$^{-4}$ provides evidence for a molecular phase.
No other sightline through the Stream to a UV-bright continuum source
has yielded an H$_2$ detection, indicating a low covering fraction for
the molecular phase, 
although there are H$_2$ detections in the Leading Arm 
(Sembach et al. 2001) and Bridge (Lehner 2002),
and recently Murray et al. (2015) reported HCO$^+$ absorption 
from the leading edge of the Bridge.
The cool H I, H$_2$, and HCO$^+$ detections demonstrate the multi-phase nature
of the gas in the Stream. This is also revealed by the many forms of ionized
gas that are present, as we discuss in Section \ref{plasma}.

The length of the known Stream has grown with time as more diffuse
gas clouds have been discovered.
Nidever et al. (2010) confirmed that the numerous small clouds in
the ``northern extension region'' near the Galactic plane
(defined by Braun \& Thilker 2004)
indeed belong to the Stream. This means 
that the Stream and Leading Arm together 
cover over 200$^\circ$ in length on the sky.

\subsubsection{Chemical abundance patterns}
\label{abundances}

The presence of metals in the Stream was first demonstrated by Songaila (1981),
who used optical spectroscopy of the background quasar Fairall~9 to detect
Ca II and Na I absorption at the Stream's velocity.
These lines trace cool clumps of dense neutral gas,
and have since been reported in other 
Stream directions (West et al. 1985, Fox et al 2013, Richter et al. 2013). 
As Ca II and Na I fall in the optical, they can be observed at higher spectral 
resolution and higher sensitivity than is possible with UV lines that 
require space-based observations. However, both Ca II and Na I are 
non-dominant (trace) ionization stages, and so these lines are of limited use 
for metallicity measurements. Nonetheless, observing them
helps to reveal the clumpiness and velocity structure of the cool gas.

Measuring the metallicity of the Stream requires a combination of 
metal-line and H I observations.
UV absorption-line spectra provide the metal-line column 
densities and 21 cm emission-line spectra provide the H I column densities.
Until the installation of the Cosmic Origins Spectrograph
(COS) on the {\it Hubble Space Telescope} in 2009, only a very 
limited number of AGN lying behind the Stream were bright 
enough for UV spectroscopic observations (Savage et al. 2000). 
One of the few exceptions is the QSO Fairall 9, which lies behind the
head of the Stream close to the LMC. Lu et al. (1994) measured two
constraints on the Stream's metallicity along this sightline: 
a silicon abundance (Si/H)$\ge$0.2 solar and 
a sulfur abundance (S/H)$\le$0.3 solar.
A metallicity measurement (rather than a limit) in the Fairall 9 
sightline was made by Gibson et al. (2000), 
who found (S/H)=0.28$^{+0.17}_{-0.10}$ solar,
consistent with a Magellanic origin, but this measurement does not 
distinguish between LMC and SMC scenarios. 
In the tip of the Stream (farthest from the Magellanic Clouds),
Fox et al (2010) measured 
(O/H)=0.10$\pm$0.02 solar, the first indication of a metallicity
\emph{below} the current day SMC abundance of $\approx$0.2 solar.

Recent work on the chemical abundances of the Stream 
using \emph{HST}/COS has found that the two principal filaments 
show different chemical abundances. 
The filament shown by Nidever et al (2008) 
to connect kinematically to the LMC indeed shows 
LMC-like abundances 
($\approx$0.5 solar; Gibson et al. 2000; Richter et al. 2013).
The other filament shows SMC-like abundances in multiple directions
($\approx$0.1 solar toward RBS~144, NGC~7714, and NGC~7469; Fox et al. 2013).
These COS measurements were made using the relatively undepleted elements sulfur
or oxygen, via the (S II/H I) or (O I/H I) ratios, 
so are not thought to be 
significantly affected by dust depletion. They were also made in directions
of high H I column $N$(H I)$>$10$^{20}$\sqcm, which are predominantly neutral 
and self-shielding, so ionization corrections are expected to be small.
The significance of these measurements is the confirmation that \emph{both 
Magellanic Clouds contributed gas to the Stream}.

The finding that the measured abundance of the ``SMC filament'' is 0.1 solar, 
\emph{lower} than the mean current-day SMC abundance of $\approx$0.2 solar 
(Russell \& Dopita 1992) could have several explanations.
First, it could indicate that the gas was stripped in the 
past, when the mean SMC metallicity was lower. Indeed
the SMC's age-metallicity relation 
(Pagel \& Tautvaisiene 1998; Harris \& Zaritsky 2004; Cignoni et al. 2013)
indicates that the mean SMC metallicity was $\approx$0.1 solar 
$\approx$2--2.5\,Gyr ago, in agreement with the time when tidal 
models find that the main body of the Stream was formed 
(Besla et al. 2010, see Section \ref{dynamics}).
The timescales for this model are therefore self-consistent: the gas
has the right metallicity for its tidal age.
Alternatively, the low measured abundance could indicate that the Stream
was stripped from outer, less metal-enriched regions of the SMC, 
since peripheral gas would be more loosely bound gravitationally and hence
easier to strip. However, the SMC does not show a strong 
radial chemical abundance gradient (Pagel et al. 1978, Cioni 2009), 
at least at the present time.
A final possibility is that some degree of metal mixing between the cool 
gas and the surrounding hot gas has diluted the 
Stream, lowering its initial metallicity, although this is hard to quantify,
since metal mixing between different gas phases is a poorly understood process,
and the metallicity of the hot gas phase is itself poorly known.

Further information on the chemical enrichment of the Stream is provided
by observations of the compact high-velocity cloud (CHVC) 224.0-83.4-197 
located close ($<$0.7$^\circ$ separation) to the Stream along 
the line of sight to the bright AGN Ton~S210. The CHVC has a metallicity 
consistent with 0.1 solar (Sembach et al. 2002, Richter et al. 2009),
with the latest measurement being [O/H]=$-$1.12$\pm$0.22 (Kumari et al. 2015). 
Based on the match between this cloud's metallicity and that of the
nearby filaments of the main Stream, Kumari et al. suggest the cloud may be 
an outlying fragment of the Stream. This is the only small outlying
cloud whose metallicity has been shown to match the Stream itself;
it is interesting that no obvious dilution of the gas (which would
have lowered the metallicity) has occurred.

\subsubsection{Dust content}
\label{dust}

The relative abundance of refractory elements, which deplete onto dust grains,
and volatile ones, which do not, 
provides an empirical measurement of the dust depletion level in the 
Stream. The earliest application of this technique was by Lu et al. 
(1994), who found little evidence for dust in the Fairall 9 direction 
(through the LMC filament) based on 
a measurement (Si/S)$>$0.6 solar. However, 
in the Leading Arm toward the Seyfert galaxy NGC~3783, Lu et al (1998) found 
a highly super-solar S/Fe ratio (7.6$\pm$2.2 solar), indicative
of dust because iron is depleted onto dust grains but sulfur is not.
Sembach et al (2001) came to a similar conclusion based on a 
{\it Far-Ultraviolet Spectroscopic Explorer (FUSE)} 
spectrum of the same AGN, and further concluded that the dust 
grains in the Leading Arm had been processed, with the grain mantles
modified or stripped to expose the grain cores.

In the SMC filament where the overall metallicity is 0.1 solar, 
Fox et al. (2013) found evidence for dust
in the form of sub-solar Si/S, Al/S, and Fe/S ratios, with
$\delta$(Si)$\approx$--0.6, $\delta$(Al)$\approx$--0.7, and 
$\delta$(Fe)$\approx$--0.6, 
where $\delta$(X)=[X/S]--[X/S]$_{\odot}$, i.e. the depletion relative
to sulfur.
This demonstrates that the dust grains survived the process(es)
that formed the Stream and also survive in the incident ionizing radiation 
field. In the LMC filament toward Fairall 9, 
where (S/H)=0.5 solar, Richter et al. (2013) found a low 
[$\alpha$/N] ratio of $-$0.85\,dex, indicating the gas there is chemically 
young, i.e. the $\alpha$-elements have been released by Type II supernovae
but N has not yet been released, since that requires a longer 
($\approx$250\,Myr) timescale until intermediate-mass stars evolve into 
the AGB phase. Richter et al. find 
$\delta$(Al)$\approx$--0.62, $\delta$(Si)$\approx$--0.27, and 
$\delta$(Fe)$\approx$--0.56, similar to the depletions in the SMC filament,
despite the factor of five difference in the overall metallicity 
in the two directions.

Combining the depletions with the H I column densities yields
the gas-to-dust ratios (G/D) in the Stream. These were calculated
for two sightlines by Fox et al. (2013).
In the SMC filament toward RBS144, the G/D ratio normalized to 
the local Galactic ISM value is (G/D)$_{\rm norm}$=19$^{+10}_{-6}$.
In the LMC filament toward Fairall 9, (G/D)$_{\rm norm}$=3.3$^{+0.5}_{-0.5}$,
a factor of $\approx$6 lower than in the SMC filament.
Since the LMC filament has a metallicity a factor
of $\approx$5 higher than the SMC filament,
the dust mass per unit gas mass scales almost linearly with 
the metallicity.

An independent approach for studying the dust content of the Stream
is to look for thermal far-IR emission in regions of high H I column density. 
Fong et al. (1987) conducted such a search for regions of the Stream near 
the south Galactic pole, and found no correlation with the IRAS 100-micron 
flux, indicating that the Stream has a lower dust-to-gas ratio than
interstellar gas in the MW, consistent with the UV-depletion
measurements.

\subsubsection{Ionized gas}
\label{plasma}

Two principal ionization processes operate on the gaseous Stream: 
photoionization and collisional ionization
(we ignore the potential contribution from cosmic rays). 
UV absorption-line observations 
(Lu et al. 1994, 1998; Sembach et al. 2001, 2003; 
Fox et al. 2005, 2010, 2013, 2014; Richter et al. 2013, Kumari et al. 2015)
reveal that both low-ionization and high-ionization phases of gas are present
in the Stream, with distinct kinematic properties. 
The high ions C IV, Si IV, and O VI
often show broader line widths and offset velocity centroids (by $<$20\kms)
compared to the low ions such as C II, Si II, Fe II, and Al II, but still 
show absorption in the same general regions of velocity space
(such high-ion/low-ion profile differences are a common feature
of circumgalactic environments).
This indicates that the ionized gas in the Stream is multi-phase, i.e. it 
contains regions of different gas density and temperature. 

The low-ion phase can be successfully modeled as a photoionized plasma
illuminated by Lyman-continuum photons. These photons have multiple sources, 
including the MW disk, the Magellanic Clouds, and the extragalactic UV 
background. The MW radiation itself has contributions from several 
components, including O-B stars, planetary nebulae, and 
potentially the central supermassive black hole. 
The relative contribution from those sources has been modeled by 
Bregman \& Harrington (1986),
Bland-Hawthorn et al. (1999, 2013) and Fox et al. (2005, 2014).
Calculating these radiation fields allows photoionization models to be 
generated for the metal-line column densities measured in the Stream, 
which in turn constrains the physical conditions in the gas, such as 
gas density, pressure, and ionized-hydrogen column density. 
The H$^+$ column can be 
combined with constraints on the cross-sectional area of the Stream to yield 
the ionized gas mass. The photoionization models are therefore crucial
for understanding the overall properties of the low-ion phase.
Representative values from photoionization models for the
gas density, thermal pressure, and line-of-sight cloud size are
log ($n$/cm$^{-3}$)=$-$1.8 to $-$2.2, $P/k$=30--250 cm$^{-3}\,K$, 
$l_{los}$=0.7--20\,kpc (Fox et al. 2014).

The high-ion phase is traced by absorption in O VI (Sembach et al. 2003) and 
C IV and Si IV (Lu et al. 1994, Fox et al. 2010, Richter et al. 2013,
Kumari et al. 2015) and
is thought to be collisionally ionized.
The main argument that collisional ionization is favored over 
photoionization is that the photoionization models that explain 
the low-ion column densities
under-produce the observed high-ion columns by orders of magnitude.
The underproduction is most severe for O VI,
where the average Stream column is $\langle$log $N$(O VI)$\rangle$=14.10 
(Sembach et al. 2003). 
Another way to state the problem is 
that there are not enough highly energetic 
($E\!>\!113.6$\,eV) photons to ionize the observed amounts of O VI 
within reasonable pathlengths 
(they require cloud sizes of $\sim$100s of kpc, which are incompatible 
with having to fit into a small volume of the Galactic halo).
However, photoionization may be an important contributor to the C IV and Si IV,
which can be created by photons of 47.9\,eV and 33.5\,eV, respectively,
and such photons are expected from the spectra of hot stars as well as
the (harder) extragalactic ionizing background.

Since there is clearly (a) cool gas in the Stream and (b) an external medium
with which the Stream interacts, it is often argued that
the high ions exist in the collisionally-ionized boundary layers between the 
H I gas and the hot coronal plasma.
However, it is unclear what the energy transport mechanism in those layers is;
they could be conductively heated, turbulently mixed, and/or shocked. 
Each of these scenarios has been explored in simulations that track the
gas ionization under certain physical conditions 
(e.g. Gnat \& Sternberg 2009, Gnat et al. 2010) or in full hydrodynamic
simulations (e.g. Kwak \& Shelton 2010, Kwak et al. 2011). 
Each model can be tested by comparing the high-ion column density ratios it 
predicts with those observed in the Stream via UV observations, such as 
$N$(C IV)/$N$(O VI) and $N$(Si IV)/$N$(C IV).
For one outer-Stream sight line (toward the QSO HE~0226--4110),
Fox et al. (2005) concluded that
either turbulent mixing or conductive heating
is a viable explanation for the high ions in the Stream.
This explanation is also favored for other HVCs in the Galactic halo
(Collins et al. 2005; Ganguly et al. 2005), with turbulent mixing layers
favored over conductive interfaces in the large dataset presented by 
Wakker et al. 2012.
Nigra et al. (2012) also favor turbulent mixing over conductive heating 
based on the size of an extended diffuse layer seen around a small clump of 
the Stream in deep 21\,cm observations.

\ha\ emission traces the warm ionized gas in the Stream in a complementary 
manner to the low-ion UV absorption lines.
\ha\ is a recombination line whose intensity scales as density squared, whereas
the strength of UV absorption scales linearly with the density.
Thus the \ha\ emission preferentially traces the densest regions of plasma 
in the Stream.
The first measurements of \ha\ emission from the Stream
(Weiner \& Williams 1996) found intensities in the range 0.20--0.37 
Rayleighs\footnote{1 Rayleigh is 
10$^6$/4$\pi$\,photons\,cm$^{-2}$\,sr$^{-1}$\,s$^{-1}$, or equivalently 
1.7$\cdot$10$^{-6}$\,erg\,cm$^{-2}$\,sr$^{-1}$\,s$^{-1}$ at \ha.}. 
These authors interpreted this emission as the signature of an 
extended gaseous halo that is ram pressure-stripping the cool gas 
in the Stream.
Putman et al. (2003a) found that the \ha\ emission is more variable, with 
intensities between 0.10 and 0.41 Rayleighs. Although the \ha\ intensity 
of other (less distant) HVCs has been used as an approximate distance indicator
(Bland-Hawthorn et al. 1999, 2002; the \ha\ intensity scales linearly 
with the incident Lyman continuum flux, which declines with distance from 
the Galaxy), the variability of the \ha\ from the
Stream challenges its use as a distance indicator.
\ha\ emission was detected from several small clumps (in the MS IV region) 
by Yagi et al. (2012), who favored a shock-cascade origin for the \ha\ 
(Bland-Hawthorn et al. 2007, Tepper-Garcia et al. 2015) 
since these clumps lie at the leading edge of a downstream cloud.
A more recent survey of 17 Stream directions with the Wisconsin \ha\ Mapper 
(WHAM; Barger et al. 2015, in prep.) finds that the \ha\ emission often extends
several degrees off the edges of the H I contours, as if the \ha\ traces
the extended surfaces of the Stream's neutral clouds and filaments.

An alternative possibility is that the \ha\ emission from the Stream was 
photo-excited
by a burst of Galactic Center (GC) activity (such as a Seyfert flare) 
$\sim$2\,Gyr ago (Bland-Hawthorn et al. 2013). In this scenario the Stream
was ionized by GC Lyman-continuum photons and is now recombining and 
emitting \ha. The $\sim$2\,Gyr timescale is plausible since (a) it is similar 
to the estimated age of the Fermi Bubbles that surround the GC and are powered 
by some form of GC activity (Su et al. 2010), and (b)
the nuclear wind has been dated to $\sim$2--4\,Gyr based on velocity
measurements from UV absorption lines (Fox et al. 2015).
The variability of the observed
\ha\ from the Stream makes it difficult to search for enhanced 
emission in a cone below the south Galactic pole, which is a signature of this 
GC flare model, but such enhanced ionization might be detected by a sufficient 
number of UV absorption-line observations of suitably-placed QSOs.

In addition to photoionization from external radiation fields, 
{\it in situ} photons emitted by the cooling hot plasma seen in O VI absorption
can also photoionize H$\alpha$ (see appendix in Wakker et al. 2012). 
Shock ionization can also contribute to the \ha\ emission from the Stream 
(Bland-Hawthorn et al. 2007; Tepper-Garcia et al. 2015).
In conclusion, multiple ionization processes may contribute to 
the Stream's \ha\ emission, and full treatments will require hydrodynamical 
simulations with radiative transport of internal and external radiation fields.

\subsection{The Magellanic Bridge} 
\label{bridge}

The Magellanic Bridge of gas connecting the Magellanic Clouds was discovered
in neutral hydrogen emission by Hindman et al. (1963), before the discovery
of the Stream.
Earlier work by Kerr et al. (1954) had shown the H I in the Clouds
to be more extended than their stars, but had not detected the diffuse
bridge in between the two galaxies.
The Bridge has historically been treated as a different object 
than the Stream, 
and this distinction makes sense for a number of reasons: 
the two are spatially separate on the sky, 
a stellar population exists in the Bridge 
(Irwin et al. 1990, Demers \& Irwin 1991, Harris 2007, Bagheri et al. 2013,
No\"el et al. 2013, Skowron et al. 2014)
but has not been found in the Stream 
(Recillas-Cruz 1982, Br\"uck \& Hawkins 1983, 
Guhathakurta \& Reitzel 1998, though see Belokurov \& Koposov 2015),
and the two structures are likely to have been formed at different times 
(Besla et al. 2012).
However, given the recent progress in understanding the dynamics of
the entire Magellanic System (Section \ref{dynamics}), the Bridge and Stream
can also be seen as separate components of a larger structure.

Like the Stream, the (gaseous) Bridge is well characterized.
It has an H I gas mass of 1.84$\cdot$10$^8$\msun\
(for $d$=55\,kpc; Br\"uns et al. 2005).
Profile analyses show that the Bridge contains two principal
components, at +214 and +234 km/s (McGee \& Newton 1986), with
other components also present that can be connected to the SMC and LMC.
Metallicity measurements based on UV absorption-line studies
of two embedded hot stars (DI~1388 and DGIK~975)
find abundances of [Z/H]=$-$1.02$\pm$0.07 
and --1.7$<$[Z/H]$<$--0.9 (Lehner 2002; Lehner et al. 2008).
Along a line-of-sight toward a QSO lying behind the Bridge, 
Misawa et al. (2009) find --1.0$<$[Z/H]$<$--0.5. 
The fact that all these values are close to the (current-day)
SMC metallicity and not the LMC metallicity suggests that the 
Bridge was formed from material formerly in the SMC. 
The column densities of H I in the Bridge are high enough 
that Ca II absorption,
which typically traces dense neutral gas, has been detected 
(Smoker et al. 2000, 2005).
In the models of Besla et al. (2012), one potential scenario for Bridge 
formation was a direct collision between the SMC and LMC $\sim$100--300\,Myr 
ago. These models predict a metallicity gradient along the Bridge
(increasing toward the LMC) owing to a contribution from LMC gas.
This prediction could be tested by future observations.

\begin{figure}
\includegraphics[width=5in]{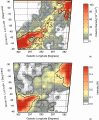}  
\caption{The Magellanic Bridge in H I (top) and \ha\ (bottom),
using observations from GASS and WHAM, respectively (Barger et al. 2013).
The SMC tail is at the lower left and the LMC is at the upper right.
The GASS data are described in McClure-Griffiths et al. (2009).}
\label{mbfig}
\end{figure}

Warm ionized gas in the Bridge was surveyed in H$\alpha$ by Barger et al. 
(2013) using the Wisconsin \ha\ Mapper (WHAM) telescope 
(see \textbf{Figure \ref{mbfig}}). They found a warm H II mass of 
(0.7--1.7)$\cdot$10$^8$\msun\ in a region where M(H I)=3.3$\cdot$10$^8$\msun.
Their analysis shows that ionizing radiation from the extragalactic background
and the MW is insufficient to explain the observed \ha\ flux; under a 
pure photo-ionization model (with no shock ionization of \ha), they 
use the observed \ha\ intensities to derive an average 
escape fraction of ionizing photons of $<$4.0\% for the LMC and $<$5.5\% for 
the SMC. Detections of cool H I emission (Kobulnicky \& Dickey 1999),
H$_2$ absorption (Lehner 2002), and CO emission (Muller et al. 2003, 
Mizuno et al. 2006) show that the Bridge (like the Stream) is a multi-phase 
structure.

In their discovery of a stellar population in the Bridge,
Irwin et al. (1990) found blue (young) main sequence stars between
the SMC wing and the Western halo of the LMC. This young Bridge
has been shown by Skowron et al. (2014) to form a continuous connection 
between the two Clouds. In addition, 
intermediate-age and old stellar populations exist
(Bagheri et al. 2013; No\"el et al. 2013).
The origin models for the Magellanic System discussed in Section 
\ref{streamorig}
need to account for this basic observation that the Bridge contains
stellar populations whereas the Stream does not (at detectable surface
density).

\subsection{The Leading Arm}
\label{la}

The Leading Arm is the counterpart to the Stream on the opposite 
side of the Magellanic Clouds. 
Parts of the Leading Arm are visible in the original H I data presented by
Mathewson et al. (1974), but its physical connection to the Clouds was not 
established until the work of Putman et al. (1998).
The study of Lu et al. (1998), who measured a sulfur abundance of 
(S/H)=0.25$\pm$0.07 solar (similar to the SMC metallicity)
in the Leading Arm toward the AGN NGC~3783, was also key in demonstrating 
the Magellanic origin of the Leading Arm.
This remains the best metallicity measurement
in the Leading Arm to date.

The Leading Arm extends $\approx$60$^\circ$ from the Clouds across the Galactic 
plane into the northern hemisphere where it
disintegrates into a large number of small cloudlets (see Figure 1). 
For et al. (2014) and Venzmer et al. (2012) recently demonstrated that
four principal Leading Arm sub-structures exist. 
The fact that the Leading Arm ``leads'' the orbits of the 
Magellanic Clouds in their motion around the Galaxy
is an important constraint on its origin, since only
tidal forces (and not ram-pressure stripping) are able to explain the presence
of gas ahead of the Clouds' trajectory.

The Leading Arm deviates from the great circle defined by the Stream by 
$\approx$60$^\circ$ (Putman et al. 1998). 
It has an H I mass of 3.0$\cdot$10$^7$\msun\ 
(Br\"uns et al. 2005) and shows high positive velocities in the LSR frame 
(from $\approx$+180\kms\ to +270\kms).
Unlike the Stream, the Leading Arm has a fairly constant column density
profile along its length (Nidever et al. 2010).
Venzmer et al. (2012) have analyzed the role of drag forces in creating the
Leading Arm structures seen in the GASS H I data. They find an inverse correlation 
between Galactic Standard of Rest velocities and Magellanic Longitudes 
(so the gas farthest from the Clouds is moving slower), and that around one 
quarter of Leading Arm clouds show head-tail
morphologies that indicate an interaction with an external medium
(see also McClure-Griffiths et al. 2008).

Casetti-Dinescu et al. (2014) recently detected 19 young massive stars 
in the Leading Arm (with ages $\approx$50--200\,Myr), 
providing evidence for {\it in-situ} star formation, since there has 
not been enough time for O-stars stripped from the Clouds to reach their 
current location before evolving off the main sequence
(an LMC star would have had to travel at $\approx$10$^4$\kms, which is 
clearly unrealistic).
This shows that dense, star-forming molecular gas exists within the Leading Arm.
Casetti-Dinescu et al. argue that the star formation may be triggered by
the interaction with the Galactic corona. 
Such triggered star formation has not been seen in the
Stream, perhaps because it is further away and interacting with a 
more tenuous halo medium.
The evidence for triggered star formation in the Leading Arm 
is consistent with the high-level of disintegration and sub-structure observed
in the Leading Arm (Putman et al. 1998, Venzmer et al. 2012).

\subsection{Total gas mass and inflow rate onto the Galactic halo} 
\label{totalmass}

The neutral gas mass of the Stream and the other
components of the Magellanic System is well established from 
21\,cm surveys (Br\"uns et al. 2005). The four dense concentration MS I--IV
account for $M$(H I)=1.25$\cdot$10$^8$\msun\ assuming $d$=55\,kpc,
the mean of the SMC and LMC distance,
although this is likely a lower limit on the Stream's mass,
since models find it could extend to $d$=100 kpc or beyond 
(Besla et al. 2010, 2012).
A further $M$(H I)=1.49$\cdot$10$^8$\msun\ exists in 
the ``Interface region'' defined by Br\"uns et al
between the Magellanic Clouds and the rest of the Stream.
In this review we treat the Interface region as part of the Stream, 
since the filaments of the Stream can be traced through
it back to the Clouds, and so it is clearly part of the same structure.
The Bridge accounts for 1.84$\cdot$10$^8$\msun\ and the Leading Arm 
3.0$\cdot$10$^7$\msun\ of H I (Br\"uns et al. 2005). Together, the H I 
in the Magellanic System (not including the H I within the LMC and SMC 
themselves) amounts to 4.87$\cdot$10$^8$\msun, comparable to the H I mass
within each of the LMC and SMC (see Table 1).

Determining the ionized gas mass is more challenging, since
one must use metal lines as proxies for the ionized hydrogen atoms
and make corrections for ionization and metallicity,
which are both non-trivial. Furthermore, 
the metal lines are measured in pencil-beam directions
and the covering fraction of ionized gas on the sky is larger than 
that of neutral gas, so a correction must be made for the cross-section.
Fox et al. (2014) constrained the Stream's gas mass using
photoionization models based on the Si II and Si III column densities 
measured in 19 extragalactic directions together with calculations
of the radiation field emerging from the MW and the Magellanic Clouds. 
They report a total gas mass for the Magellanic System
(Stream, Bridge, and Leading Arm) of $\approx$2.0$\cdot$10$^9$\msun, of
which $\approx$25\% is in neutral gas and $\approx$75\% is in 
ionized gas (including warm ionized and highly ionized components). 
Given that the LMC contains 4.0$\cdot$10$^8$\msun\ of interstellar H I and the
SMC contains 4.4$\cdot$10$^8$\msun\ at the current time
(Staveley-Smith et al. 1997, Br\"uns et al. 2005), 
and that ionized gas is not thought to be a 
dominant contributor to the Clouds' global ISM mass budgets 
(most \ha\ emission in the LMC and SMC comes from supershells, not diffuse gas;
Kennicutt et al. 1995), 
this leads to the conclusion that the
Stream, Bridge, and Leading Arm together contain 
over twice as much gas as the current-day Magellanic Clouds.
That is, if all the observed Magellanic gas (neutral and ionized) used 
to exist in the Clouds, then \emph{most of the original gas content of the 
Magellanic Clouds has been stripped}.
Thus while the Magellanic Clouds are often referred to 
as ``gas-rich'' satellites, we are observing them at a time when
they may have lost most of their initial gas.
As a note of caution, some of the ionized gas in the Stream may
trace condensing hot material from the surrounding medium 
(Sect. \ref{ms-fate}), in which case the gas did not all originate from
within the Clouds.

Dividing the Stream's gas mass by its inflow time onto the MW
gives its (time-averaged) inflow rate.
The inflow time is $\sim$0.5--1.0\,Gyr for
$d$=50--100\,kpc and its average 
Galactocentric inflow velocity of 100\kms\ 
(Mathewson et al. 1977, Mathewson 1985). 
This gives a mass inflow rate of $\sim$4--7\smy\ (Fox et al. 2014).
This is larger than the inflow rate represented by all other Galactic
HVCs, which is in the range 0.08--1.4\smy\ depending on assumptions about
distance and ionization (Shull et al. 2009, Lehner \& Howk 2011, 
Putman et al. 2012).
It is also larger than the current Galactic star formation rate, which
is in the range $\approx$1--2\smy\ 
(Chomiuk \& Povich 2011, Robitaille \& Whitney 2010).
{\it Therefore the Stream is bringing in fuel at a rate sufficient
to elevate the future Galactic star formation rate}, and at a rate higher
than that of any other observable infalling fuel source. 
The key question is whether the Stream will
survive to reach the disk, evaporate into the hot halo, or
even seed the cooling of hot halo gas and accrete more
mass as it comes in. We will return to this question in Section \ref{ms-fate}.

\section{ORIGIN OF THE MAGELLANIC STREAM}
\label{streamorig} 

The Stream has historically been explained as the outcome of 
two competing scenarios:\\

{\bf The Tidal Model.} In this scenario the Stream is the outcome either of the
tidal interactions between the LMC and the MW, as first suggested 
by Fujimoto \& Sofue (1976) and Lin \& Lynden-Bell (1977), 
or as material pulled out from the SMC 
by the tidal force exerted by the LMC and controlled by 
gravitational force of the MW, a model that traces back to 
Murai \& Fujimoto (1980) and more recently to Guglielmo et al. (2014). 
The tidal model envisions that a close 
encounter between the Magellanic Clouds occurred 
when they were separated by a distance of 2--3 kpc, with the dwarfs 
assumed to be a binary pair on short periodic orbits ($\sim$2 Gyr) 
around the MW. 
Several observations supported the tidal origin of the Stream, 
including the presence of the 
Bridge of gas and stars and the discovery of the Leading Arm. 

{\bf The Drag Model}. In this scenario the Stream is 
created by ram-pressure stripping as the Clouds pass through
an external medium, which could be either the tenuous coronal gas in the
Galactic halo (Meurer et al. 1985; Moore \& Davis 1994)
or alternatively the denser gas of the Galactic disk.
The latter would apply if the Clouds crossed
the outer disk at some point in their past orbit. 
Later studies of these effects showed that a combination of 
tidal and hydrodynamical interactions can produce a close
approximation of the Stream under the assumption that 
the Clouds had multiple close passages with our Galaxy 
(Gardiner \& Noguchi 1996; Mastropietro et al. 2005; Connors et al. 2006). \\

However, recent \emph{HST} proper-motion measurements of
the Clouds and developments in current models of galaxy formation 
have provided crucial new information regarding the Stream's history and 
origin. 
Because the new measurements suggest that
the Clouds have either completed one orbit around the MW or may
even be at first perigalacticon, a revised interpretation of the origin 
of the Stream may be required.
Numerous observed features of the Stream, Bridge, and Leading Arm
need to be accounted for by origin models and simulations.
 
\begin{itemize}
\item The spatial extension of the Stream and the Leading Arm, which 
together stretch to over 200$^\circ$ 
on the sky, possibly crossing the Galactic disk in two locations, at 
the Leading Arm and at the Stream tip. 

\item The Stream presents a filamentary structure (Wakker 2001; Putman
  et al. 2003b) with at least two main filaments (Nidever et al. 2008), 
  one with metallicity consistent with the LMC, and one more close in 
  metallicity to the SMC (Fox et al. 2013, Richter et al. 2013).

\item A dominant fraction of the Stream's mass is in the form of ionized 
 gas (Fox et al. 2014). 

\item There is lack of a stellar counterpart in the Stream, 
  despite the plentiful
  amount of gas in neutral and ionized form.
  This absence of stars in the Stream  prevents 
  distance measurements. However, recently 19 young stars {\it in situ} 
  have been discovered in
  the Leading Arm, providing an approximate estimate to that region of 
  $\sim 21$ kpc (Casetti-Dinescu et al. 2014).  
\end{itemize}

It has been argued
that the Stream is a young feature (1--2 Gyr; Besla et al. 2007),
which provides a constraint on origin models. 
The arguments for a young Stream are twofold, but have attached caveats.
First, simulations of the survivability of high-velocity clouds moving
in a hot external medium (Heitsch \& Putman 2009; Joung et al. 2012) 
find that clouds evaporate on timescales of hundreds of Myr to $\sim$1 Gyr. 
However, the cloud lifetime depends on the HVC mass: these simulations
only considered HVCs with H I masses $<$10$^{4.5}$\msun, whereas
the largest Stream clumps are more massive (and longer lasting). 
Second, the Stream exhibits high \ha\ emission 
(see Section \ref{plasma}), which may indicate that its gas
is being ablated away on a 100–-200 Myr timescale
(Bland-Hawthorn et al. 2007).
However, the origin of the \ha\ emission is unclear: it may
be shock-ionized and/or photoionized, so the \ha\ intensity 
does not provide a clean diagnostic on age.
Therefore, there are no clear observational constraints on the Stream's age.

Despite the observational constraints listed above and numerous 
theoretical efforts and observational constraints,  
the origin of the Stream and Leading Arm remain uncertain. 
Recent $N$-body and hydrodynamic simulations 
(Besla et al. 2010, 2012; Diaz \& Bekki 2011b, 2012) have
questioned  the standard picture of Stream formation in which the
Clouds traveled on a quasi-periodic orbit around the MW.  
These new models posit that the origin of the Stream was caused by the mutual 
tidal interaction between the Clouds \emph{before} they were 
accreted by the MW (Besla et al. 2010; 2012) and on a bound orbit around the
MW (Diaz \& Bekki 2011b,2012). The Galactic potential governs 
the orbits of our neighboring dwarfs and
therefore causes the trail of the Stream to be amplified across the sky once
the pair of dwarf galaxies fell in. 
The recent proper-motion estimates of the Clouds are consistent with one or
two passages of the Clouds around the MW -- a third passage is less likely --
making it difficult to discriminate among these tidal scenarios.

\subsection{First-passage (unbound) scenarios}
\label{firstpassage}
In the first-passage scenario the LMC and SMC are just past their first
pericentric passage (Besla et al. 2007). In this model
the LMC first entered the virial radius of the MW within the past 
1--4 Gyr and has not yet completed an orbit.
This model was updated by Besla et al. (2012) to explore the morphology of the
Stream produced from a head-on collision between the Clouds, specifically 
by the SMC moving in a highly eccentric orbit around the LMC, far from the 
MW potential. 
Highly eccentric orbits are cosmologically motivated as 
suggested by studies of satellite galaxies in current models of galaxy 
formation, but were never assumed in early studies on the formation of 
the Stream. While the high eccentricity of the orbit prevents the Clouds 
from merging in these new models, it requires the SMC to 
spend most of its time at apocenter (at $\sim$80--100 kpc away from the LMC).
At the first collision with the LMC the Stream forms as material tidally
removed from the SMC, a process that does not require any interaction 
with the MW potential 
(and this is consistent with the Stream's low metallicity). 
The Clouds are assumed to be an 
interacting pair for a significant fraction of the Hubble time, and
only recently fell into the MW potential (about 2 Gyr ago). 

There is compelling evidence that the LMC and SMC had
a recent and close encounter. First, the Magellanic Bridge
of gas and stars appears to be a tidal feature
(see e.g. Besla et al. 2012). 
Second, the LMC and SMC proper motions imply at least one
collision within the past 500 Myr, as a direct result of
the relative orientation of their three-dimensional velocity vectors
(Kallivayalil et al. 2013).  
Third, the distribution of OB stars in the MCs and 
near the Bridge is also consistent with a recent exchange of material, 
200 Myr ago (Casetti-Dinescu et al. 2013).

This picture is plausible within
the current $\Lambda$CDM paradigm where halos 
at all scales build up their mass hierarchically. 
Interacting dwarf galaxies are therefore cosmologically expected both 
in isolation and on their orbits around more massive galaxies.
When two disky galaxies pass by one another, in this case the SMC and LMC
disks, their mutual gravitational tidal forces distort the disks of stars and 
gas resulting in the formation of tails and bridges of material (Toomre \&
Toomre 1972). It has been shown that the efficiency of the process 
depends particularly on the inclinations of the disks relative to the 
orbit plane. Therefore, the mechanism is more efficient for co-planar 
and prograde encounters than for retrograde ones 
(Toomre \& Toomre 1972; D'Onghia et al. 2010; \L okas et al. 2015). 

\begin{figure}
\label{simstream}
\includegraphics[width=5in]{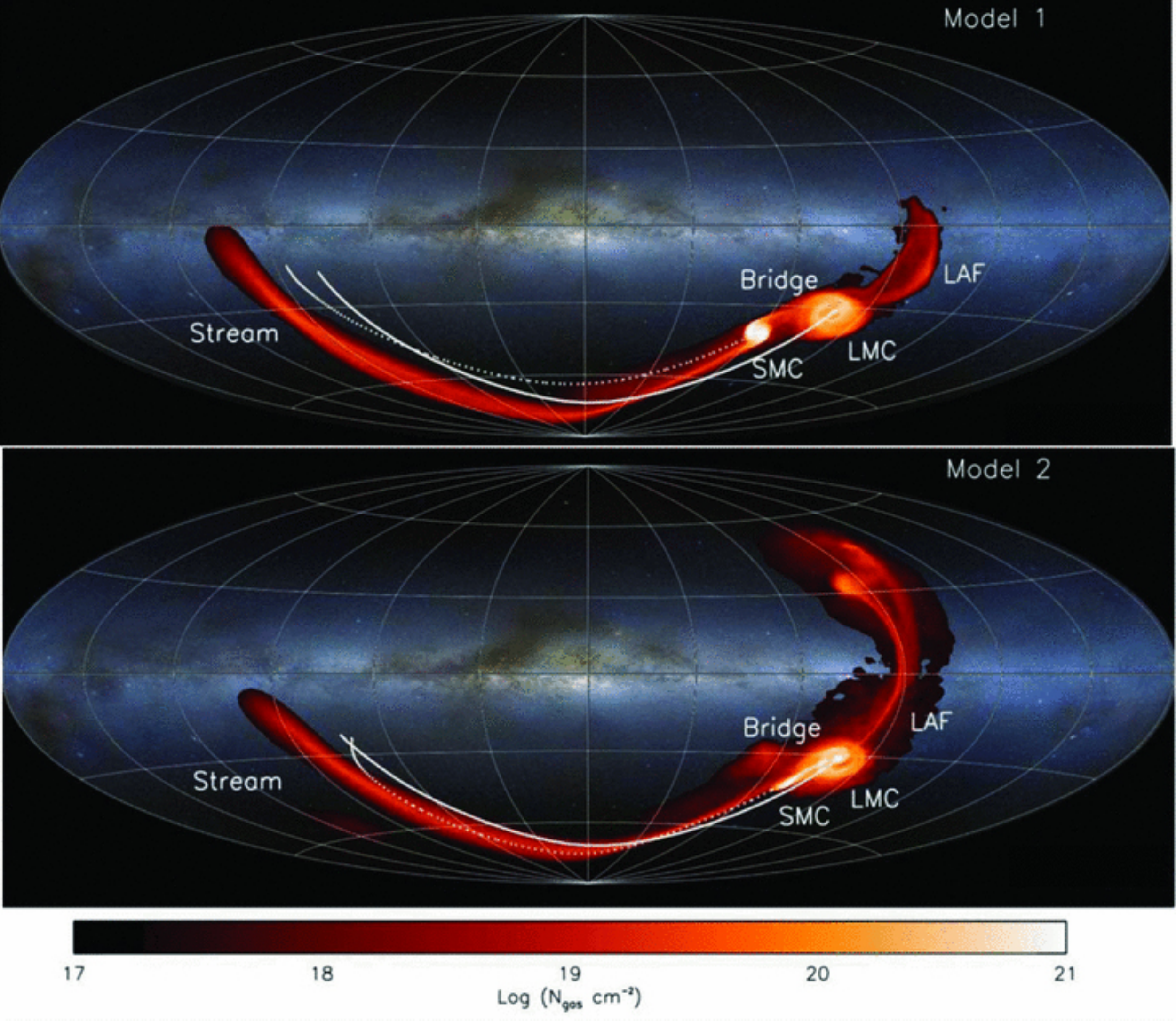}
\caption{Gas projection in the simulated Magellanic System (red scale) 
plotted over an optical image of the MW. The models reproduce the
length of the Stream and Leading Arm, as well as the Bridge connecting
the Clouds. 
In the top panel (Model 1), the SMC has completed two passages around the LMC.
In the lower panel (Model 2), it has completed three.
Credit: Besla et al. (2012), with background image from Mellinger (2009).}
\end{figure}

The Besla et al. (2012) first-passage models reproduce many 
observed features of the Stream (see {\bf Figure 4}) 
in particular:

\begin{itemize} 
\item [1)] The location of the Stream across the sky is well approximated; 
\item [2)] The resulting line-of-sight velocities for the simulated
 Stream agree with the data; 
\item [3)] The simulated H I column densities range from 
10$^{18}$--$10^{21}$ cm$^{-2}$ in agreement with the data, 
although the exact column density gradient along the length of the Stream 
is not reproduced (see Figure 8 in Besla et al. 2012). 
The column density is inhomogeneous across the width of the
simulated stream: the inclusion of a realistic interaction with the 
MW halo gas might help in reproducing the bifurcated, filamentary nature 
of the observed Stream; 
\item [4)] The pronounced asymmetry between the trailing and leading
components is matched. Indeed, the Leading Arm of the simulated Stream 
is much smaller than the trailing component, as observed. This arises 
because the leading tidal arm from the SMC falls toward the LMC, 
while the trailing component stretches out to increasing distance. 
\end{itemize}

Some difficulties are still experienced in the first-passage scenario, 
that call for additional physics to be eventually included in the models. 
In particular, the H I column  density is one
tenth of that observed (Besla et al. 2012). The
tidal model underestimates the Stream's gas mass by
a factor of $\approx$10--100 when accounting for the huge (but not precisely
constrained) amount of ionized gas. 

Another difficulty with the first passage model 
is that one filament of the Stream originates 
in the LMC (Nidever et al. 2008; Richter et al. 2013). 
The fact that the LMC filament traces back to the starburst
region 30 Doradus strongly suggests that star formation is linked to
its origin.
This is difficult to reconcile with the
tidal model where the Stream gas is formed purely
by direct collision between the Clouds. 

A recent study explored the implications of the 
ram-pressure stripping experienced by the LMC moving at high speed 
through the Galactic halo gas at its first passage with the MW,  
employing both analytic prescriptions and 
full 3-dimensional hydrodynamic simulations (Salem et al. 2015). 
While this study shows that it is unlikely that material ram-pressure stripped
from the LMC can account for more than a tiny 
percentage of the total mass of the Magellanic
Stream, it does show that gas located 
to the North-West of the LMC's disk is enriched in metallicity by the 
contribution of gas ram-pressure stripped by the LMC.

The position of the Leading Arm in the models is inconsistent with its observed
location. There is some evidence that the Leading Arm has already 
reached the Galactic disk: McClure-Griffiths et al. (2008) report an 
interaction between the Leading Arm cloud HVC 306-2+230 and the disk, 
which gives a kinematic distance to 
this part of the Leading Arm as $\approx 21$ kpc from the Sun, consistent
with the distance to the newly-discovered stars in the Leading Arm 
(Casetti-Dinescu et al. 2014). 
This distance is close to that predicted by the Connors et al. (2006) and 
Yoshizawa \& Noguchi (2003) models, that advocate for at least one past passage
of the Clouds with the MW.  

It should be stressed that the first-infall 
scenario naturally reproduces the Bridge,
through a high-speed encounter with the
SMC passing through the center of the LMC. However,
one question that remains puzzling and needs clarification 
concerns the metallicity and age of the Bridge. 
The Bridge has a measured gas-phase metallicity of only 0.1 solar 
(Lehner et al. 2008, Misawa et al. 2009), lower than the current-day
abundances of both Clouds, and its age might indicate an earlier formation than
200--300 Myr ago, when the models posit it was formed. 

\subsection{Multiple-passage (bound) scenarios}
\label{secondpassage}
In multiple-passage models where the Clouds follow a bound orbit, 
there is a clear difference in the
role played by the MW in the origin of the Stream. 
The key feature of these models is that there is time for tidal 
interactions to occur between the Clouds.
If the Clouds are on a bound orbit, the
MW seems to govern the binary action of the LMC--SMC pair, guiding
them into a recently formed short-period orbit (Diaz \& Bekki 2011b). 

The Diaz \& Bekki model posits that the LMC-SMC system only recently formed a
binary pair, within the last $\sim$2 Gyr. 
In this scenario, the LMC and SMC may have originally formed
as independent satellites of the MW, separated
by large distances (Bekki et al. 2004). Their 
orbital evolution through the Galactic halo 
gradually brought them closer together until the LMC was able to
capture the SMC into its orbit and form a tightly bound binary pair. 
While it is not clear how likely such a satellite capture is in models of
galaxy formation, a bound orbit with the Clouds currently being on their
second perigalacticon is not ruled out by the current measurements 
of the proper-motion of the Clouds. 
Interestingly, the model seems to reproduce the on-sky bifurcation 
of the Stream's two filaments better than the first-passage model,
suggesting that a bound association with the 
MW favors the formation of such bifurcation.

A study of the global SFH of the LMC suggests that the LMC and SMC might have 
had coupled episodes of star formation, one occurring $\sim$2 Gyr ago 
and one $\sim$500 Myr ago (Harris \& Zaritsky 2009). It is unclear whether 
the timings of the two close LMC--SMC interactions in the first-passage 
and second-passage models are consistent with these burst epochs. 
The central theme of these tidal scenarios where the LMC and SMC are a strongly
interacting pair is that the Stream originates from material pulled out of the
Clouds. Because the gravitational field of the LMC will act in the same 
manner on gas and stars, we expect a tail of stars pulled out
from the SMC into the Stream, especially in the case of a head-on collision. 
Unfortunately,
this stream of stars still awaits discovery; searches for a stellar
Stream have been unsuccessful (Recillas-Cruz 1982, Br\"uck \& Hawkins 1983, 
Guhathakurta \& Reitzel 1998), although on the opposite (Eastern) side of LMC,
a (possibly unrelated) stellar stream 10 kpc long has recently been reported
by Mackey et al. (2015; see also Belokurov \& Koposov 2015). 
While the 19 recently-discovered young stars
in the Leading Arm (Casetti-Dinescu et al. 2014)
appear to have formed {\it in-situ}, it is unclear whether
this star formation was triggered by
the interaction with the Galactic disk or halo.
Future metallicity measurements of the Leading Arm gas near these stars 
will give insights on their origin.

\subsection{The role of ram-pressure stripping} 
\label{origother}
A pure ram-pressure-stripping origin for the Stream has been
explored by many authors (Meurer et al. 1985, Moore \& Davis 1994,
Heller \& Rohlfs 1994, Murali 2000, Mastropietro et al. 2005, 
Diaz \& Bekki 2011a). 
Such models naturally explain the lack of stars in the
trailing Stream, but do not explain the existence 
of the Magellanic Bridge connecting the LMC and
SMC or the origin of the Leading Arm (which leads
the entire system). It is thus clear that tidal interactions between the
Clouds must have played a role in the formation of the Magellanic System.

However, the detailed morphology of the Stream and the Leading Arm 
prove the complexity of the Magellanic System. 
Several simultaneous physical processes are needed to describe it.
Recently, the  Parkes  Galactic All-Sky Survey (GASS; McClure-
Griffiths et al. 2009) 
was used to identify the locations where the filaments 
cross each other and to explore the twisting filamentary system
at higher resolution (Hammer et al. 2015).  In an attempt to reproduce
these two filamentary structures using hydrodynamic
simulations, this study assumes that the MCs were following 
parallel orbits along the observed  Stream. Thus, the Stream  
from the South Galactic Pole to the current location of the Clouds would 
be generated by a direct collision between the Clouds and the interaction 
with the MW hot gas. Leaving aside the question of how the Clouds are 
accreted into the Galactic halo, the model shows that the pair 
started to respond to the MW hot halo gas at a large distance from the 
MW and then had a mutual collision with a pericenter of $\approx$3 kpc 
$\approx$250 Myr ago. The outcome is two prominent ram-pressure tails 
trailing behind each Magellanic Cloud.
The assumption of the Clouds traveling on parallel orbits and their recent
collision produces the twin, intertwined filaments, as observed.
The recent collision also produces a Magellanic Bridge. 
The resulting Stream from these models is shown in {\bf Figure 5} 
It would be interesting to assess the likelihood of such a
orbital configuration for the Clouds in cosmological simulations. 

\begin{figure}
\label{hammer}
\includegraphics[width=5in]{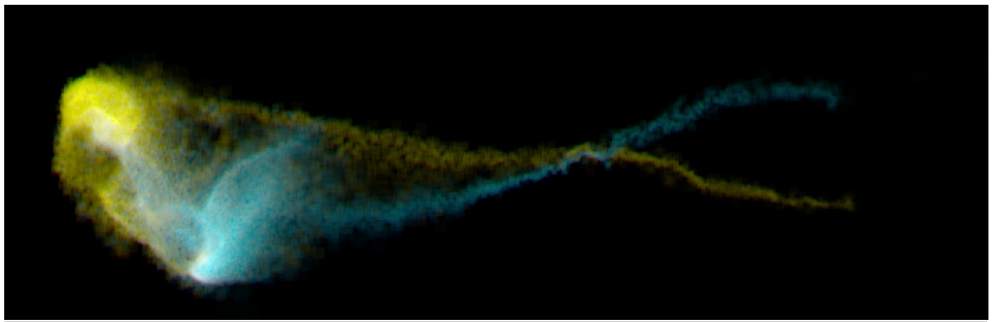}
\caption{Simulation of the Stream that reproduces
its spatial extent and filamentary nature.
In this model the Magellanic Clouds collided
$\approx$200--300 Myr ago and the two filaments of the Stream are
formed by the ram pressure exerted by a hot MW corona with a gas density 
$\approx$10$^{-4}$ cm$^{-3}$. Gas from the LMC is shown in yellow; gas from
the SMC is shown in blue. Credit: Hammer et al. (2015).}
\end{figure}

\subsection{Other scenarios}
An alternative to the canonical tidal and ram-pressure scenarios envisions
a \emph{primordial origin} for the Stream (Peebles \& Tully 2013). 
According to this model, the Stream originated from tidal interactions at 
high redshift between the young LMC and the MW, which was its nearest 
massive neighbor at the time.
This picture traces back to the model introduced by 
Fujimoto \& Sofue (1976) and Lin \& Lynden-Bell (1977), but 
applied at high redshift under cosmological initial conditions
and solving the equation of motion for the Local Group.  
A first test of the viability of this model would be to
check whether a primordial scenario can account for the Stream's length,
width, and later interactions with the SMC, that we
know at this point to be responsible of the Bridge.

It is worth mentioning that if the Clouds fell into the MW halo as part of a
Magellanic Group (see Sect. \ref{group}) then they would naturally have had 
high-speed encounters in the past and the tidal processes 
invoked to account for the origin of the Stream would still apply, 
but it would be interesting to check with hydrodynamic simulations. 

\section{FATE OF THE MAGELLANIC STREAM}
\label{ms-fate}

The Stream is clearly a substantial gas reservoir, with 
a mass-flow rate high enough to potentially elevate the future star formation 
rate of the MW (Sect. \ref{totalmass}).
However, one of the biggest open questions on the Stream
is whether it will survive its inflow passage and reach the Galactic 
disk. The hydrodynamic interaction with the hot Galactic corona may 
dissipate the Stream and replenish the corona with new material.
Several lines of evidence indicate that such an interaction is already underway.

First, there is clear evidence for small-scale structure in H I,
in the form of fragmentation and cloudlets observed in many locations around
the main filaments of the Stream, as if the Stream is disintegrating
as it plunges into the halo 
(Mirabel et al. 1979, Mirabel 1981,
Putman et al. 1998, 2003b, Wakker et al. 2002, Stanimirovi\'c 
et al. 2002, 2008, Westmeier \& Koribalski 2008, For et al. 2014).
Furthermore, Putman et al. (2003b) and For et al. (2014) found 
many of these clouds show head-tail 
cometary morphologies, with a dense core and a diffuse tail. 
These morphologies are indicative of cloud/corona interactions 
(Br\"uns et al. 2000, Quilis \& Moore 2001, Peek et al. 2007), in which
the head indicates the direction of the cloud's motion and
the tail contains material swept into the cloud's wake.

Second, UV absorption-line studies have shown that the Stream 
contains a high-ion phase traced by the Si IV, C IV, and O VI doublets
(Sembach et al. 2003, Fox et al. 2005, 2010, 2014, Richter et al. 2013, 
Kumari et al. 2015), as discussed in Section 3.3.
This phase appears to be collisionally ionized 
at a characteristic temperatures of a few times 10$^5$\,K
and will arise at the boundary layers between
the cool Stream and the hot corona. The high ions can be thought
of as a signature of this interaction.

Third, hydrodynamical simulations have explored the lifetime of cool gas 
clouds against disruptive encounters with the coronal gas
(notably Kelvin-Helmholtz instabilities), and find that 
for mass ranges appropriate for HVCs, the timescales are relatively short.
Heitsch \& Putman (2009) found that for HVCs with H I
masses $<$10$^{4.5}$\msun\ the evaporation timescales are $<$100\,Myr.
The less massive the cloud (or cloudlet), the shorter the disruption time.
Similar disruption timescales are found by Bland-Hawthorn et al. 2007 and 
Joung et al. 2012. 
Magnetic fields, if present, may stabilize the clouds and 
increase their lifetime against instability.

Ultimately, whether the Stream survives its inflow
passage to the MW depends on two timescales. If the gas disruption time 
is less than the inflow time ($\sim$0.5--1.0 Gyr), the Stream will not
survive. However, 
the appropriate disruption timescale to adopt is not straightforward,
because of the fragmentation of the Stream into structures
on a range of spatial scales. The larger, coherent structures
such as the principal filaments are more likely to survive, but 
the smaller cloudlets ($<$10$^{4.5}$\msun) will not.
Furthermore, the presence of a magnetic field, which has been detected
at the $\sim$6 $\mu$G level
in the Leading Arm (McClure-Griffiths et al. 2010), may stabilize
cool clouds against evaporation.

This picture becomes more complicated once the condensation of coronal 
gas is considered. When a cool cloud interacts with the hot halo,
the two phases mix at the cloud boundary, leading to the 
evaporation of cool gas and the condensation of hot gas.
Depending on which process dominates, the mass of the cloud can either
shrink or grow with time. These processes have been explored in the 
context of other HVCs in the Galactic halo 
(Marinacci et al. 2010, Fraternali et al. 2013, 2015; Marasco et al 2013).
The balance of evaporation vs. condensation 
depend on the metallicity of the corona,
since the condensation is driven by cooling, which scales with metallicity.
It appears unlikely that condensation would dominate in a case with 0.1 solar
metallicity, as is appropriate for most of the Stream, since the cooling rate
is so low in this regime. 

\section{KINEMATICS AND DYNAMICS OF THE MAGELLANIC SYSTEM}
\label{dynamics}

\subsection{Kinematics of the LMC}
\label{lmckin}
The dynamics, structure, and star formation history of the LMC
have long been interpreted in the context of its proximity to both the MW and
its dwarf-galaxy companion SMC. 
This has allowed for detailed and thorough analyses of its structure 
and kinematics, with much work in the
last decade based on measuring
the photometry and kinematics of large samples of individual stars, and 
deep star counts to measure the true extent of the LMC. 
Our knowledge of the internal kinematics of the LMC
has improved over the last two decades,
with different observational tracers covering
different components of the Magellanic System.
This includes kinematic measurements of the 
H I gas (Kim et al. 1998; Staveley-Smith et al. 2003), 
star clusters (Grocholski et al. 2006), 
and stars (Prevot et al. 1985; Massey \& Olsen 2003, 
Kunkel et al. 1997; van der Marel et al. 2002, 
Olsen \& Massey 2007; Olsen et al. 2011).  

The rotation curves of galaxies are based on line-of-sight velocities
and so are one-dimensional.
For most galaxies, proper-motion measurements are not possible
within current observational capabilities, so three-dimensional
kinematic information is unavailable, but the Magellanic Clouds are
an exception. The recent LMC proper-motion measurements led to the 
first measurement of the large-scale rotation field in three dimensions 
(van der Marel \& Kallivayalil 2014). 
These measurements are fundamental to our understanding of the
dynamics and mass of our largest dwarf companion. 
The proper-motion measurements imply an amplitude for
the LMC rotation curve of $v_{0, PM}=76.1 \pm 7.6$ km/s, obtained with a
magnitude-limited sample of stars with mixed populations. 
This measurement falls in
the middle between the measurements of the line-of-sight velocities of old and
young stellar populations with velocities $v_{0, LOS}=55.2 \pm 10.3$ km/s and
$v_{0, LOS}=89.3 \pm 18.8$ km/s, respectively (van der Marel \&
Kallivayalil 2014). The major uncertainties on these measurements 
are due to the inclination of the LMC. 

The rotation curve of the LMC peaks at $v_{cir}=91.7 \pm 18.8$ km/s,
which implies an enclosed mass M$_{\rm LMC}=(1.7\pm0.7)\cdot10^{10}$\msun\ 
within a radius of 8.7 kpc (see Table \ref{msys}). 
The total dynamical mass of the LMC extends beyond this
radius but is hard to determine. Given the fact that the LMC is in tidal
interaction with the MW, its dark halo might be tidally truncated. An estimate
of the inferred tidal radius is 22.3$\pm$5.2 kpc 
(van der Marel \& Kallivayalil 2014). 

Furthermore, through the study of the LMC proper motion rotation field it
has been possible to address the location of the stellar dynamical 
center of the LMC, and whether it coincides with the H I dynamical center.
Different measurements of the LMC's stellar center using 
different components or techniques are not spatially coincident, indicating 
that there is not a single well-defined center (van der Marel 2001; 
Cole et al. 2005). The proper-motion rotation fields recently inferred 
addressed this question. These measurements 
indicate that the dynamical center inferred by proper motion agrees with 
the H I dynamical center and the offset with the line-of-sight is greatly 
reduced. Uncertainties coming from the assumptions made in the model 
to infer the line-of-sight velocities may also introduce an offset between the
dynamical center and the large line-of-sight velocity-field. 
In particular, the structure of the LMC turned out to be
more complicated than a flat circular rotating disk (van der Marel \& 
Kallivayalil 2014). 

An ongoing area of research related to the structure of the LMC concerns 
its stellar bar. 
Despite a wealth of data there is still great uncertainty 
concerning the bar's nature. The work of van der Marel (2001) 
indicates that the bar is off-centered from the dynamical center of the LMC
and that it resides within a large stellar disk. 
Later work proposed that this feature might not be a bar, but 
instead a triaxial stellar bulge residing in
a thick disk (Zaritsky 2004). Interestingly, 
the bar is not present in the comprehensive maps of the H I
distribution and kinematics of the LMC (Kim et al. 1998, 
Staveley-Smith et al. 2003), despite the fact that the feature is clearly 
visible in the stellar distribution of the LMC. 
The OGLE III survey (Udalski et al. 2008) 
has been used to argue that the LMC’s bar resides in the plane of the disk 
(Subramaniam \& Subramanian
2009). Earlier work described the bar as being an unvirialized structure 
that is offset from the rest of the disk as a result of the LMC's 
interaction with the SMC (Zhao \& Evans 2004). Recent
numerical experiments of collisions between the SMC and LMC dwarf galaxies
supported these findings (Besla et al. 2012, Yozin \& Bekki 2014). 
However, $N$-body simulations of interacting dwarf galaxies with properties 
similar to the Clouds show that the dynamical center and the bar center always 
coincide, but the photometric center 
is mismatched with the dynamical center due to 
tidally-induced distortions of the stellar disk (Pardy et al. 2015, in prep.).
If this is the case it would be interesting to place 
the LMC dynamical center on the
bar instead of the HI dynamical center. This would slightly change the north
component of the proper motions: $\mu_N$, perhaps reducing the offset 
between the LMC's orbit and the position of the Stream. 

\subsection{Kinematics of the SMC}
\label{smckin}

The intermediate-age and old stellar populations of the SMC, traced by 
carbon stars (e.g. Kunkel et al. 2000) and planetary nebulae (Dopita et
al. 1985) show an absence of rotation. However, the young stellar population, 
traced by H I gas, shows a velocity gradient (Staveley-Smith et al. 1997). 

The kinematics of the different components of the SMC are
consistent with the picture of the SMC originally having been a disk-like 
dwarf galaxy before experiencing recent strong tidal interactions with the 
LMC. While there is no evidence of tell-tale tidal tails, 
the velocity field of the central region of the SMC 
is consistent with a recent tidal encounter with the LMC.
When a spinning system like a disky dwarf galaxy encounters a more
massive system like the LMC or the MW, it will lose mass by tidal stripping
and will undergo a morphological transformation, with angular momentum
being removed after the encounter (Mayer et al. 2001; D'Onghia et al. 2009). 
However, even after several encounters a velocity gradient is still expected
to be present in the central regions of the SMC if gas and stars are 
rotationally supported, which seems to be consistent with current data. 
The SMC rotation curve inferred by H I data shows a rise up to a
distance of 3 kpc from the center and reaches a maximum of 60 km/s,  
corresponding to an enclosed total mass of 2.4$\cdot$10$^9$\msun,
within that 3 kpc radius (Stanimirovi\'c et al. 2004). Detailed
three-dimensional proper-motion measurements like the ones inferred for the
LMC are needed to better constrain the properties of the SMC, especially in
light of its role in the formation of the Stream. 

\subsection{Proper motions and orbital history of the Magellanic Clouds}
\label{propermotions}

The unusual situation of having a pair of Magellanic galaxies so close 
to a large spiral like the MW results in interactions 
in the LMC-SMC-MW system not seen in isolated Magellanic pairs. 
This situation is so uncommon that recent studies based on cosmological
simulations and observations have concluded that less
than 10\% of MW-like galaxies are expected to
host satellites with properties similar to the Clouds (Busha et al.
2010; Boylan-Kolchin et al. 2011, Tollerud et al. 2011).

A description of the global dynamics of the Magellanic System is 
needed before processes like star formation, 
chemical evolution, and kinematics in the LMC and SMC can be fully understood.
Conversely, these processes can give information on parameters of the 
interactions. Thus the interacting Magellanic System is a key probe into 
many aspects of our Galaxy, from its mass to its halo environment, 
as well as being a doorway to understanding 
the formation and evolution of galaxies in general.
 
High-precision proper-motion measurements of LMC stars based
on \emph{HST} data taken in the last decade
(see Section \ref{pmhst}; Kallivayalil et el. 2006a,b, 2013;
Piatek et al. 2008) indicate that the LMC has a higher tangential velocity 
than previously thought (the latest measurement is $v_{tan} \sim 314$ km/s; 
Kallivayalil et al. 2013). 
A high tangential velocity for the LMC means that the centrifugal
force is larger than the gravitational force acting
on the LMC, thus the LMC is moving toward larger radii 
and so must be now at perigalacticon. The high tangential velocity
raises questions as to whether the 
Clouds are actually bound to each other or even to the MW as a whole.

Early models assumed that the Clouds' orbits were slowly 
decaying into the MW halo potential by dynamical friction 
(Murai \& Fujimoto 1980, hereafter MF80; Davies \& Wright 1977;
Lin \& Lynden-Bell 1977, 1982; Gardiner et al.
1994; Heller \& Rohlfs 1994; Moore \& Davis 1994; Lin et al. 1995; 
Gardiner \& Noguchi 1996; Bekki \& Chiba 2005, 2007; Yoshizawa \&
Noguchi 2003; Connors et al. 2004, 2006; Mastropietro
et al. 2005).
Because of the large uncertainties in previous proper-motion measurements,
the orbital parameters were chosen to best reproduce the properties 
of the Stream under the assumption 
that the LMC and SMC form a binary system that has been in a slowly 
decaying orbit around the MW for nearly a Hubble time. 

In the next section, we discuss the proper-motion measurements 
of the Clouds based on the three-epoch \emph{HST} data and discuss 
how these measurements affect the orbital history of the LMC 
and SMC and hence the origin of the Stream.

The proper motion is defined in the west ($\mu_W$) and north ($\mu_N$)
directions as the variation in time of the right ascension $\alpha$ and 
declination $\delta$ in the sky:

\begin{equation}
\mu_W=-(d\alpha /dt) cos \delta, \hspace{1cm} \mu_N=-d\delta /dt
\end{equation}

\noindent
The two components of the proper motion control the orbital 
parameters of the LMC. The combined west and north components 
determine the tangential velocity
of the LMC and thereby sets the orbital period, number of
pericentric passages, apogalactic distance, and the stability 
of the SMC--LMC binary system.
The north component controls the location of the orbit when projected on 
the plane of the sky. 

\begin{table}
\caption{LMC and SMC Proper Motion Measurements}
\label{pms} 
\begin{tabular}{lll llr} 
\hline
Study Method &    LMC     &       &   SMC  &    &  Refs.\\

             &   $\mu_W$  &    $\mu_N$ &  $\mu_W$ &   $\mu_N$ & \\ 
             &   (mas/yr) &    (mas/yr) &  (mas/yr) &  (mas/yr) & \\
\hline
MS Model (G94) & $-$1.72           &   0.12            &  ...   & ... & (1,2)\\
MS Model (HR94)& $-$2.0       &   0.16  &  ...   & ... & (3)\\
Ground-based+Hipparcos$^{a}$ & $-$1.68$\pm$0.16  & 0.34$\pm0.16$ & ... & ... & (4) \\
\emph{HST} two-epoch (K1,K2) & $-$2.03$\pm$0.08  &   0.44$\pm$0.05   &
$-$1.16$\pm$0.18  & $-$1.17$\pm0.18$ & (5)\\
\emph{HST} two-epoch (P08) & $-$1.56$\pm$0.036 &  0.435$\pm$0.036 &
$-$0.754$\pm$0.061 & $-$1.252$\pm$0.058 & (6) \\
2.5m du Pont   & $-$1.72$\pm0.13$  &   0.50$\pm$0.15   &
$-$0.93$\pm$0.14 & $-$1.25$\pm$0.11 & (7) \\
Southern Proper Motion  & $-$1.89$\pm$0.27  &   0.39$\pm$0.27   &  $-$0.98$\pm$0.30   &
$-$1.10$\pm$0.29 & (8) \\
\emph{HST} three-epoch (K3) & $-$1.910$\pm$0.02 &   0.229$\pm0.047$ &
$-$0.772$\pm$0.063 & $-$1.117$\pm$0.061 & (9)\\
\hline
\end{tabular}
\begin{tabnote}
References: (1) Gardiner et al. 1994; (2) Gardiner \& Noguchi 1996; (3)
Heller \& Rohlfs 1994; (4) van der Marel et al. 2002; (5) Kallivayalil et al
2006a; 2006b; (6) Piatek et al. 2008; (7) Costa et al. 2011; (8) Vieira et
al. 2011; (9) Kallivayalil et al. 2013.\\
$^{a}$ Value weighted average of the ground-based and {\it Hipparcos} 
measurements by Kroupa et al. 1994; Jones et al. 1994; Kroupa \& Bastian 1997;  
Pedreros et al. 2002; Drake et al. 2001. \\
\end{tabnote}
\end{table}

\subsection{Proper motions from the \emph{Hubble Space Telescope}}
\label{pmhst}

The advent of high-precision proper-motion measurements of the Clouds 
using \emph{HST} data (Kallivayalil et al. 2006a, 2006b, 2013, 
hereafter K1, K2, K3; Piatek et al. 2008, hereafter P08) and a more
physically motivated modeling of the Magellanic Clouds (Besla et al. 2007) 
have revolutionized our view of Magellanic dynamics. 
Following earlier results based on two epochs (K1, K2),
a third epoch of \emph{HST} data of ten
QSOs behind the LMC and three behind the
SMC has been recently analyzed (K3). 
The combined data give a 7-year baseline of proper-motion measurements for 
both Clouds. 
A comparison between the latest proper-motion measurements 
and the earlier measurements is given in {\bf Table \ref{pms}}. 
{\bf Table \ref{3epoch}} lists the 3D velocity in the Galactocentric restframe 
of both the LMC and SMC, and the relative velocity of the SMC with 
respect to the LMC, as reported in K3.

The LMC proper-motion measurements obtained from three epochs of
\emph{HST} data (K3) provided the following results and implications:

\begin{itemize}
\item The three-dimensional speed of the LMC is lower 
(by $\sim$57 km/s) 
than measured from the two-epoch measurements, 
owing to the decrease in the west component 
of the proper motion and the new measurement of the solar velocity.
For a MW mass of 1$\cdot$10$^{12}$\msun\
these measurements imply that the LMC might be on a parabolic orbit, 
with the LMC just past perigalacticon (K3). 
Alternatively, if the mass of the MW 
is higher at $\approx$2$\cdot$10$^{12}$\msun, the LMC will be on a 
more eccentric orbit with a period of $\approx$6 Gyr, and currently 
at its second pericentric passage (Besla et al. 2007, Piatek et al. 2008, 
Shattow \& Loeb 2009, K3). 

\item 
The current measurement of the north component of the proper motion
$\mu_N=0.229\pm0.047$ (Table \ref{pms}) indicates 
that the LMC's orbit slightly deviates from the position of the Stream on the 
sky. This modest offset implies that the LMC orbit is  not a close tracer 
of the Stream. 
The assumption that the Stream is aligned with the past orbit of the
Clouds is one of the assertions of the tidal-stripping scenario 
for the Stream's origin.
The SMC proper motion measurement suggest that in the past the SMC
crossed the location of Stream, an orbital solution that agrees with 
the picture of the Stream being torn off the Clouds by their mutual
tidal interactions (GN96; R\r{u}\v{z}i\v{c}ka et al. 2009, 2010; 
Diaz \& Bekki 2011b; Besla et al. 2012). 

\item The observed relative velocity between the Clouds indicates that 
the radial component $v_{rad}$ of the LMC-SMC system
is larger than the tangential component $v_{tan}$, 
suggesting that the Clouds are on an eccentric orbit.
\end{itemize}

The interpretation of the orbital history of the Clouds  
based on the proper-motion measurements crucially depends on 
the circular velocity inferred for our Galaxy and how it 
compares to the tangential velocity of the LMC. 

An increase of a factor of two in the virial mass of the MW makes a difference
(Piatek et al. 2008; Shattow \& Loeb 2009), which is certainly
within the current uncertainty in virial mass estimates for the MW
(e.g. Smith et al. 2007, Li \& White 2008, Reid et al. 2009). 
In fact the escape speed at $r$=50 kpc
from a NFW halo (Navarro et al. 1997) with a virial 
mass\footnote{The virial mass, M$_{200}$ is defined as the mass
contained within r$_{200}$, the radius of a sphere of mean density 200 times 
the critical density for closure, $\rho_{crit}$=3H$^{2}$/8$\pi G$. 
This choice defines 
implicitly the halo virial radius r$_{200}$ and its virial velocity V$_{200}$.} 
M$_{200}=2 \cdot 10^{12}$\msun\
is of the order of 500 km/s, which would indicate that the Clouds are 
bound to the Galaxy despite their high speed (Sales et al. 2011).

Additional uncertainties in the 
measurements of the MW's circular velocity affect the 
measurement of the proper motions of the LMC, because this is measured 
relative to the Solar System that orbits the Galaxy.
The rotational velocity of the Sun is therefore needed in 
order to transform to the 
Galactocentric frame (van der Marel et al. 2002). 
Although the recent $\sim$14$\pm6$\% increase in the MW circular 
velocity relative to the International Astronomical Union (IAU) standard 
of 220 km/s is included in the analysis of the proper-motion based on the 
third epoch of HST data (K3), the measurement of the solar proper motion 
is still debated.
 
\begin{table}
\caption{Three-epoch \emph{HST} Measurements: Galactocentric Velocities}
\label{3epoch}
\begin{tabular}{lll llll} 
\hline
Galaxy & $v_{X}$       & $v_{Y}$       & $v_{Z}$      & $v_{tot}$     & $v_{rad}$   & $v_{tan}$\\
       & (km/s) & (km/s) & (km/s) & (km/s) & (km/s) & (km/s)\\
\hline
LMC$^{*}$ & $-$57$\pm$13   & $-$226$\pm$15  & 221$\pm$19 & 321$\pm$24 & 64$\pm$7 & 314$\pm$24\\       
SMC$^{*}$ & 19$\pm$18    & $-$153$\pm$21  & 153$\pm$17 & 217$\pm$26 & $-$11$\pm$5 & 217$\pm$26\\
SMC--LMC$^{*}$ & 76$\pm$22   &  73$\pm$26 & $-$68$\pm$25 & 128$\pm$32 & 112$\pm$32 & 61$\pm$16\\
\hline
\end{tabular}
\begin{tabnote}
Credit: Kallivayalil et al. 2013\\
$^{*}$Measurements based on the velocities used to correct for solar
  reflex motion: the improved McMillan (2011) value of $v_0$=239$\pm$5 km/s
  and the solar peculiar velocity reported by Sch\"onrich et al. (2010). 
\end{tabnote}
\end{table}

\subsection{Orbital history of the LMC: parabolic or bound?}

Given the new Galactocentric velocities derived from the three-epoch 
\emph{HST} data, the challenge is to determine whether the LMC is 
on a parabolic or on a bound orbit around the MW.
This question depends on the modeling rather
than the measurements of the velocities. Thus there are good reasons to believe
that this debate will not be settled by further improving the precision of
future measurements of proper motions. On the contrary, it depends crucially on
the Galactic mass distribution, its total mass and virial radius, 
and how the Galactic potential evolves in time.

In the following we offer an example of the limitations of the models in trying
to constrain the orbital history of the Clouds. The LMC orbit has been
obtained by solving the differential
equations of motion (see Murai \& Fujimoto 1980, Besla et al. 2007, Guglielmo
et al. 2014):

\begin{equation}
\ddot{{\bf r}}=\frac{\partial}{\partial {\bf r}} \Phi_{\rm MW}(|{\bf r}|)+\frac{\mathcal{F}_{df}}{M_{\rm LMC}}
\end{equation}
\noindent
Here M$_{\rm LMC}$ is the mass of the LMC, ${\bf r}$
is its position vector, $\Phi_{\rm MW}$ is the Galactic potential, 
and $\mathcal{F}_{df}$ is the dynamical
friction term that assumes the form:\\

\begin{equation}
\mathcal{F}_{df}=-\frac{4\pi G^2 M_{\rm LMC}^2 {\rm ln}(\Lambda)
  \rho(r)}{v^2}\Big[\rm{erf}(X)-\frac{2X}{\sqrt{\pi}} \rm{exp} (-X^2)\Big]\frac{\bf
v}{v}
\end{equation}
\noindent
where $\rho(r)$ is the density of the host halo at the Galactocentric
distance of 50 kpc (the distance of the LMC), 
ln\,$\Lambda$ is the Coulomb logarithm, 
$v$ is orbital velocity of the
LMC and $X=v/\sqrt{2}\sigma$. Here, $\sigma$ is the one-dimensional 
velocity dispersion of the dark matter halo.
By imposing the current velocities and positions for the LMC, 
the differential equations of motion have been used 
to determine the position and velocities of the LMC at earlier times. 

While this approach has been useful in early works that assumed that 
the Clouds' orbit were slowly decaying into the Galactic potential by 
dynamical friction, it relies on assumptions
that require to be tested to make the orbital solution plausible.
First, the formalism assumes that 
the LMC evolves in a rigid Galactic potential, which is probably not
realistic. Second, the orbit of the LMC decays by dynamical
friction and by tidal mass-loss. However, the dynamical friction term depends
on the mass of the LMC which is tidally stripped by time, a situation that 
requires full $N$-body simulations to be captured. 

\begin{figure}
\label{k3orbits}
\includegraphics[width=5.20in]{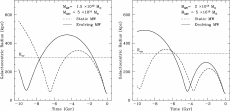}
\caption{Evolution of LMC orbits with time in the MW potential for cases
with M$_{\rm MW}$=1.5$\cdot$10$^{12}$\msun\ (left) and 2.0$\cdot$10$^{12}$\msun\ 
(right). The mass of the LMC is assumed to be
M$_{\rm LMC}=5\cdot$10$^{12}$\msun\ in both models. 
In each case, the two lines show the cases of a static (fixed MW mass; dashed) 
and evolving (increasing MW mass; solid) Galactic potential.
An evolving MW potential gives a longer orbit period for the LMC than the 
case of a fixed MW mass.
Credit: Kallivayalil et al. (2013).}
\end{figure}

{\bf Figure 6} 
shows that the LMC orbit period becomes longer if the 
Cloud feel the MW potential that evolves in time as compared to orbits 
computed in a static Galactic potential. 
Increasing the mass of MW models results in shorter LMC orbital periods.
Gomez et al. (2015) argue that a massive LMC falling into the MW can 
perturb the Galactic potential. The 
result is a dynamical drag on the LMC motion that tends to shorten its 
orbit period and favor a scenario where the LMC is at currently at past 
second perigalacticon.
In {\bf Figure 7} 
the orbits of the LMC (solid black lines) and SMC (red dashed lines) 
in the MW potential are displayed for 
two cases: first- and second passages. 
The first passage solution is obtained for a massive LMC 
falling into a massive MW halo, not accounting for the MW dynamical 
response to the massive LMC.
When the dynamical drag is accounted for the LMC orbit period shortens, 
leading to a second pericentric passage.    
These orbital solutions (Figures 6 and 7) 
illustrate how the orbital history depends on the assumption made on modeling, 
rather than on the proper-motion measurements. 

\begin{figure}
\label{orbitzy}
\includegraphics[width=5.0in]{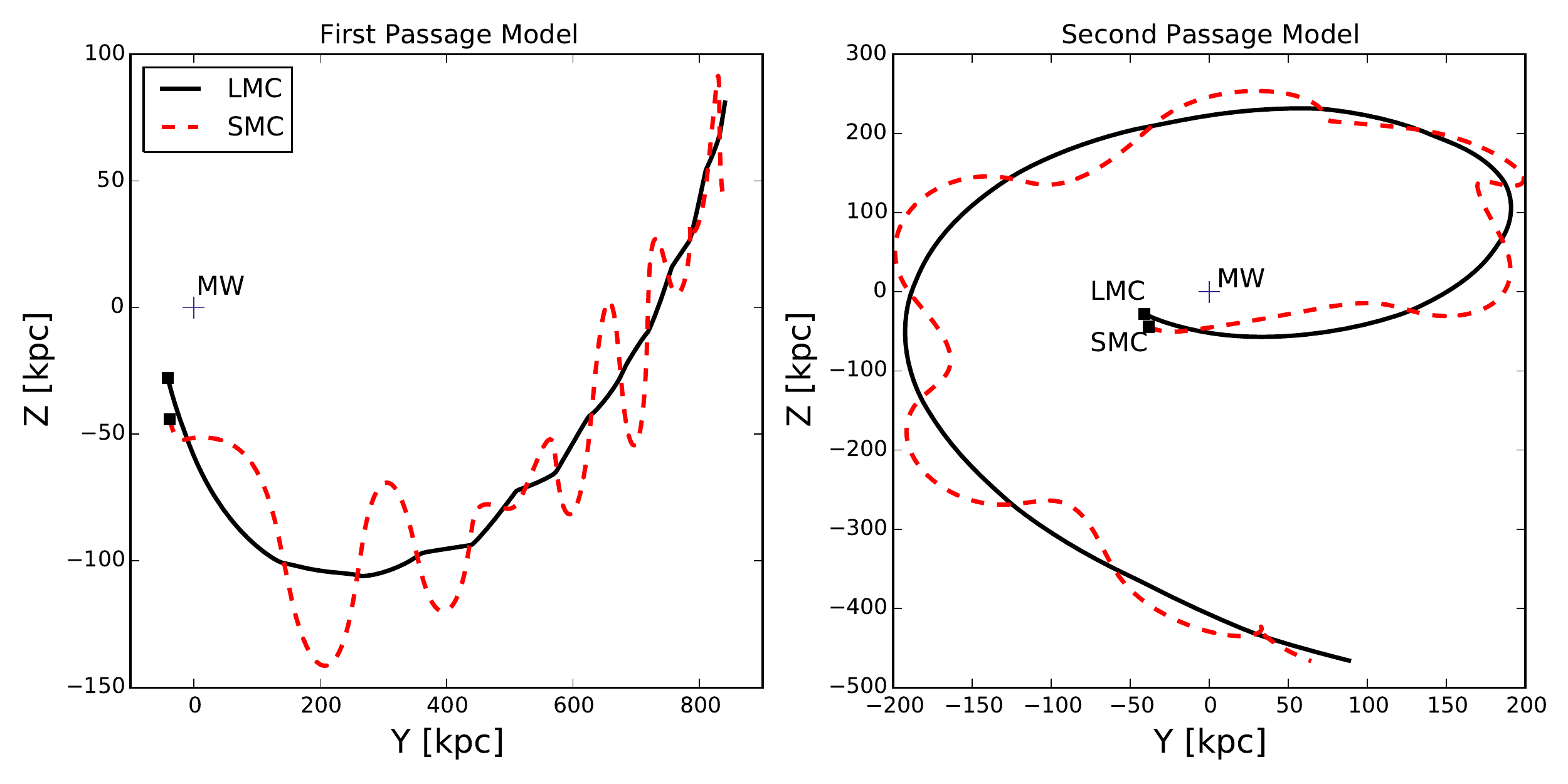}
\caption{Two distinct orbital solutions for the LMC (black curve) and 
SMC (red dashed curve) in the MW potential, illustrating the first- (left)
and second- (right) passage scenarios. 
The masses assumed for the MW and the LMC are: 
M$_{\rm MW}$=1.5$\cdot$10$^{12}$\msun\ and
M$_{\rm LMC}$=1.8$\cdot$10$^{11}$\msun. 
The effect of dynamical drag
on the motion of LMC perturbing the MW is accounted for and results 
in a shorter orbital period for the LMC,
favoring a double passage around the MW (right panel).
Credit: this work.}
\end{figure}

Another question related to the past orbital history of the Clouds 
concerns whether the LMC and SMC were and still are a bound pair. 
First-passage models where the Stream represents the debris of the 
mutual tidal interaction between the LMC/SMC {\it before} they fell 
into the MW require the Clouds to be bound at least for the last 
4 Gyr and eventually have formed the Stream in the last 2 Gyr 
(Besla et al. 2012). 
Since the relative velocity between the Clouds is $\sim$130 km/s, 
LMC masses of the order of $10^{11}$\msun\ are
needed in these models to explain how the LMC can have held on to the SMC. 
Thus these models require a low-mass MW 
($\sim$10$^{12}$\msun) and a massive LMC ($\sim$10$^{11}$\msun) to ensure
that the Clouds are a long-term binary system.  
  
In conclusion, our ability to properly model the orbital history of the 
Clouds depends on the ability to measure the potential of our Galaxy,
e.g. by accurate measurements of the rotation curve, 
and on our understanding of the
processes involved in galaxy formation, from the galaxy mass distribution, the
satellite orbits, the environment, to gas dynamics. 
Thus, the progress of cosmological
simulations of galaxy formation has the promise to better
constrain the kinematics and dynamics of the Clouds and the Stream.  
Further progress in proper motion measurements, as will be possible with the
{\it Gaia} mission, and, equally important, continued advances in the 
art of distance 
measurements,  will further tighten the constraints, and may
show us whether we really understand how dynamics is operating in our closest 
dwarf companions.
 
\subsection{The Magellanic Group}
\label{group}

The fact that the Clouds are so close to each other and embedded
in the Magellanic Bridge of neutral hydrogen advocates for their 
physical connection. 
Using a projection of the bright dwarf galaxies of the MW 
onto a map of high-velocity hydrogen, Lynden-Bell (1976) suggested that, 
like the Magellanic Clouds, other near-neighboring dwarf galaxies 
were associated with the H I Stream. This was supported
by the polar distribution of the brightest Galactic satellites, which seems
to trace the orbital path of the Clouds. 
The number of dwarf galaxies in this association has increased with 
time and initially included the following candidates: the Magellanic 
Clouds, Ursa Minor, and Draco. Sculptor and Carina would also be added 
to the association of galaxies in the Magellanic Stream, though 
their potential membership is more tentative as their orientations are
not along the Stream (Lynden-Bell \& Lynden-Bell 1995).  
The distribution of distant halo globular clusters like Pal I 
has been used to reinforce the existence of both groups 
(Kunkel 1979; Majewski 1994). 

Lynden-Bell (1976) speculated that if the LMC and SMC are physically 
associated 
then it is possible that they were once part of a larger system.
He noted that Draco and Ursa Minor appeared to be distributed
on the sky along a great circle but opposite the Magellanic Clouds, 
suggesting that they were debris torn off the so-called Greater Magellanic
Galaxy. A later study showed that the elongation of
Draco and Ursa Minor occurs along the Magellanic Stream (Lynden-Bell 1982). 
If the Clouds were at more than one pericentric passage 
around the MW, on the previous passage Sculptor and perhaps 
Draco and Ursa Minor were pulled out, 
but the next pericentric passage has been more severe with the two 
Clouds being pulled apart. 
 
More recently it has been proposed that the Magellanic Clouds were accreted
into the MW halo as the largest members of a group of dwarf
galaxies, named the Magellanic Group, that fell into the MW halo at a 
relatively recent time (D'Onghia \& Lake 2008). This scenario is motivated 
by theoretical and observational arguments. It is naturally
expected in the most popular scenario of structure formation, 
the $\Lambda$ cold dark matter (CDM) model, which predicts that the abundance
of sub-structures is self-similar. This means that an abundance 
of sub-halos is expected at all observable mass scales. Thus, if 
hundreds of satellite halos are predicted to surround the MW halo 
(Klypin et al. 1999, Moore et al. 1999), at lower mass scales dwarf-galaxy 
halos can also host sub-structures. The second motivation comes from the 
discovery of nearby associations of dwarf galaxies at distances of 1--3 Mpc 
from the Local Group (Tully et al. 2006). These systems are defined as 
associations, rather than groups, because of their low density and and 
the fact that they are loosely bound 
systems -- not in a dynamical equilibrium
-- since the crossing time is very long: 80\% of the Hubble time. 
They have masses ranging between 5$\cdot$10$^{10}$\msun\ and
1$\cdot$10$^{11}$\msun\ and typically contain a larger dwarf irregular 
galaxy and a few dwarfs comparable in mass to dwarf spheroidals, as well 
as fainter satellite galaxies. NGC~3109 is one example of such 
association, with a dwarf irregular as the largest member surrounded 
by a few fainter dwarf galaxies, including Antlia and the recently 
discovered fainter system Antlia B (Sand et al. 2015). 

The LMC may have been the largest member of such a dwarf association. 
In this scenario, when this group fell into the MW potential, 
it carried several of the bright satellites of the MW, 
such as Draco, Sculptor, Sextans, Ursa Minor, as well as Sagittarius, 
Sextans and Leo II, among others. This picture is 
supported by the evidence that in the hierarchical universe 
many satellite halos potentially hosting dwarf galaxies have been 
accreted into the MW as pairs (Sales et al. 2007, Ludlow et al. 2009; 
Klimentowski et al. 2010) or as part of multiple systems, 
as indicated by cosmological simulations  
(Lux et al. 2010, D'Onghia \& Lake 2008; 
Li \& Helmi 2008; Ludlow et al. 2009; Wetzel et al. 2015).

Observationally, the discovery of conformity positions and velocities of 
satellite pairs, such as LMC, SMC, Leo IV and Leo V (Belokurov et al. 2008), 
and of satellites close to the Sagittarius stream, e.g. Segue 1 
(Belokurov et al. 2007, Niederste-Ostholt et al. 2009), 
Bootes II (Koch et al. 2009), Segue 2 (Belokurov et al. 2009) and Segue 3
(Belokurov et al. 2010) further motivated the recent search for 
associations between satellite galaxies and streams.

Recent studies examined the orbits of the bright dwarf galaxies bound to 
the LMC in the past as part of a Magellanic Group falling into the MW at 
its first or second perigalacticon, but reported contrasting results. 
By running a 
Monte Carlo suite of models of the Magellanic Group in gravitational 
interaction with the MW and comparing the results to the available 
kinematic data for the local dwarf galaxies, Nichols et al. (2011) 
concluded that 
Draco, Sculptor, Sextans, Ursa Minor, and Sagittarius are consistent
with having fallen in along with the Magellanic System, in addition to 
cases such as Carina and Leo I, which might have a different origin.
A different study examined the infall of a Magellanic Group into the 
MW using zoom-in high-resolution $N$-body simulations of a MW-sized halo
(Sales et al. 2011).  
This study finds that Draco and Ursa Minor are unlikely 
to be accreted into the MW as part of the Magellanic Group if 
the Clouds are on their first   
perigalacticon, although there could possibly remain 
a large number of faint satellites near the Clouds.  

\begin{figure}
\includegraphics[width=5.0in]{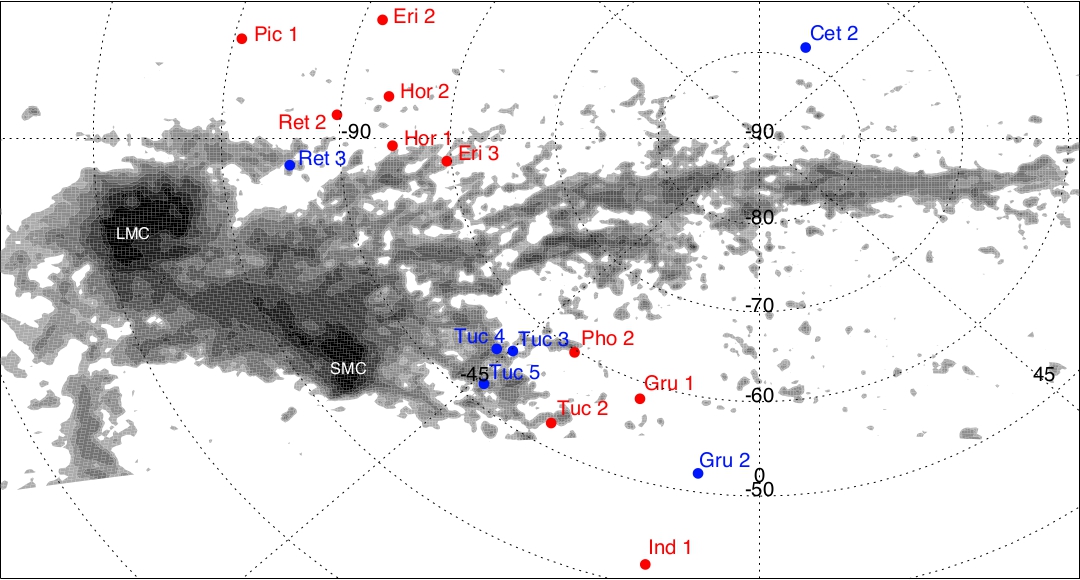} 
\caption{Location of the MW satellite galaxies with respect to the H I
Stream, including recently detected ultra-faint dwarfs found
in the DES Survey Year 1 data (red) and Year 2 data (blue). 
Credit: Vasily Belokurov, updated from Koposov et al. 
(2015) with H I data from Putman et al. (2003b).}
\label{koposov}
\end{figure}

The recent discovery of several new ultra-faint MW companions, using the data
from the Dark Energy Survey (DES), is consistent with the scenario of a
Magellanic Group (see {\bf Figure \ref{koposov}}; 
Koposov et al. 2015, Bechtol et al. 2015, Martin et al. 2015,
Westmeier et al. 2015). 
Three systems qualify to be dwarf-galaxy candidates, based on their
morphological properties, while the nature of the other six objects 
is currently uncertain. 
In particular, Reticulum 2, Horologium 1, and Eridanus 3 are 
aligned with the LMC's orbital plane and may form part of the its cortege. 
Similarly, Tucana 2, Phoenix 2 and Grus 1 appear to align with the SMC.
This suggests a picture in which at least some newly discovered 
ultra-faint dwarf galaxies,  together with some of the already 
well-known satellites, belonged to a 
loose association of dwarfs bound to the LMC and SMC, with size and mass
comparable to the dwarf associations discovered
around the Local Group. 

There is some evidence that satellite galaxies are not isotropically
distributed around massive galaxies (Bailin et al. 2008 and reference therein), 
and correlations among their spatial
distribution or possible alignments are ubiquitous. Indeed, half of the
satellite galaxies of M31 may lie in a thin, extended and
rotating plane (Ibata et al. 2013), although these findings are 
still uncertain (Phillips et al. 2015). 
Planes of satellites might be present in the Centaurus Group as
well (Tully et al. 2015).

If there was a Magellanic Group of dwarf galaxies, 
the latest measurements of the proper motion of
the LMC (K3) would suggest that it is currently 
either at its first perigalacticon or has already completed a full orbit. 

Kinematic data is inconclusive as to whether the LMC  
is on its first passage or has already completed a full orbit 
(Sect. \ref{firstpassage}). 
If at its first perigalacticon, then most of its associated satellite 
dwarf galaxies should be 
tightly clustered around its location. Although
rare, some LMC-associated systems may still be found well away
from the LMC but along the orbital path of the Magellanic Group (Sales et
al. 2011, Deason et al. 2015). 
Of the well-known MW satellites, only the SMC is unambiguously associated 
with the LMC. There is some chance that Fornax, Carina and Sculptor 
might be associated, although the available proper motions 
would indicate that the orbital planes of these three satellites might 
not be aligned with that of the Clouds. If the LMC has already 
completed one full orbit, then several additional
dwarfs qualify for association. Leo II, Leo IV and Leo V, in particular,
show strong spatial and velocity coincidence with the tidal
debris from LMC, making them candidates for past association
with the LMC, in addition to Canes Venatici II. 
These tentative associations of the known dwarf galaxies of the MW to a 
Magellanic Group will be confirmed or ruled out when their proper motions 
become available.

If a Magellanic Group fell into the MW carrying a number of satellites, 
the SMC might have played a role as well. Recent estimates of the
star-formation history of the SMC, derived using high-resolution
color-magnitude diagrams and images from the \emph{HST} 
Advanced Camera for Surveys, indicate that the SMC experienced a global peak of 
star formation activity between 5--7 Gyr ago (Cignoni et al. 2013). 
It should be noted that this result is consistent
with the sudden appearance 7--8 Gyr ago of an enhancement of 
globular clusters in the SMC (Piatti 2011). 

If a dramatic event happened in the SMC $\sim$7 Gyr ago as suggested by these 
findings, the question is whether this can constrain the past orbital history 
of the SMC.
Tsujimoto \& Bekki (2009) proposed that the SMC 
experienced a merger event 7.5 Gyr ago in a small group of gas-rich dwarfs 
with the size and mass comparable to the Clouds. While it is unclear 
whether this picture can reproduce the SMC age-metallicity relation reported
by Cignoni et al. (2013), it seems an appealing scenario to be explored with 
future numerical simulations, especially in light of the possibility that 
the Magellanic Clouds may have been accreted as a group of dwarfs instead 
of a galaxy pair. 

The discussion on the past existence of the Magellanic Group should end 
with a note of caution. Despite the fact that satellite galaxies appear 
to be distributed on the sky along a great circle rather than in a 
spherical way, the great circle has a polar orientation 
relative to the MW disk. The discovery of most of the SDSS ultra-faint 
dwarfs in the northern hemisphere might have increased the apparent 
significance of the alignment. Another important point about associations or
planes of dwarfs is that they affect the statistics. If satellites are
accreted as groups rather than individuals possible alignments are more
likely to be spotted.
Thus, the completion of the DES survey, by covering areas farther away 
from the polar structure, as well as Pan-STARRS and the next generation of 
deep wide-field surveys, will be able to characterize and fully address 
the question of the anisotropy of the distribution of MW satellites. 
Furthermore, all of these newly-discovered objects trail the LMC and SMC 
in their orbits in the sky. In the Magellanic Group scenario,
there should also be a population of satellites that leads the
LMC and SMC; such a population awaits discovery. 

\subsection{Fate of the Magellanic Clouds}
\label{cloudsfate}

A question closely related to the fate of the Stream is the fate of the
Clouds themselves, which depends on whether
they are just at their first passage or have
already completed an orbit around the MW. An old argument from Tremaine
(1976), made with the belief that the Clouds had made multiple passages 
around the MW, posited that dynamical friction must have caused substantial 
decay of the orbits of the Magellanic
Clouds over the last 10 Gyr, during which time the Galactic tidal force
at perigalacticon must have steadily increased. The LMC in this
scenario will be disrupted by the Galaxy in 2--4 Gyr, increasing the
luminosity of the Galaxy by --0.24 mag. However, these calculations did not
account for the tidal field of M31 and the new orbital solutions for the 
Clouds.
Given that the MW is approaching M31 and is expected to merge
within it in 4 Gyr (Cox \& Loeb 2008; van den Marel et al. 2012), 
this calls into question which merger will happen first. 

\section{SUMMARY AND FUTURE DIRECTIONS}

\begin{issues}[SUMMARY POINTS]
\begin{enumerate}
\item Via radio, UV, and optical observations, we now have a 
comprehensive knowledge of the morphological, chemical, 
structural, and physical properties of the Magellanic Stream, 
Bridge, and Leading Arm.

\item The recent third-epoch \emph{HST} high-precision proper motions 
measurements of the Magellanic Clouds show that the Clouds 
are either on their first passage onto the MW or on an
eccentric long orbit. This differs remarkably
from the past canonical picture in which the Clouds are our long-term
companions traveling on a quasi-periodic short orbit around the MW. 

\item As a consequence of the new orbital history of the Clouds, a new 
picture has emerged for the origin of the Stream.
Observations and theoretical arguments strongly suggest a recent collision
between the Clouds in the last 200--300 Myr that formed the 
Bridge and contributed to the Leading Arm and the trailing Stream. 
However, the Stream's total mass remains a challenge, particularly
given the large component of ionized gas present.
Its total mass is
underestimated by a factor of $\sim$10--100 in current models. In addition
the detailed observations of the filamentary structure of the Stream
indicate that the interaction with the MW gas corona and the ram pressure
stripping still play a significant role. 
Furthermore, there is clear kinematic evidence that the LMC filament 
of the Stream connects back to the SEHO region of the LMC, 
which contains the starburst region 30 Doradus.
This suggests that 
feedback from star formation contributes gas to one portion of the Stream.

\item Several basic issues such as
whether the LMC and SMC are on their first approach, or bound
to each other, or truly associated with other MW satellites, remain
at present unresolved, largely due to uncertainties in the total mass 
of the MW and LMC.

\item The Stream offers a valuable case study of
the gas accretion processes operating in an $\approx L_*$ galaxy halo,
where a merger provides gas to fuel future star formation.
It is providing copious amounts of gas to the MW's corona.
The fate of this gas depends on the 
disruptive encounter with the coronal gas, and the balance between 
evaporation and condensation.
\end{enumerate}
\end{issues}

\begin{issues}[FUTURE RESEARCH DIRECTIONS]
\begin{enumerate}

\item Ongoing surveys for a stellar component to Stream will
address the long-standing question: if it is a tidal feature, where
are its stars? The SMASH survey using the DECam camera
will map 480 square degrees of the outskirts of the Magellanic Clouds
(Nidever 2015), supplementing the 5000 square degrees covered
by the Dark Energy Surveys.

\item Sensitive future radio surveys could address questions on the extent
of the H I Stream, such as whether its tip crosses the Galactic plane.
This will constrain further whether the Cloud are at their first
or second past perigalacticon around the MW.

\item More measurements of the metallicity of the Leading Arm and how
it compares to the current-day LMC and SMC metallicities will 
constrain the formation of the Leading Arm. This necessitates finding
AGN bright enough for UV observations in Leading Arm directions with 
substantial H I column densities.

\item Further constraints on the spatial extent of the diffuse outer layers 
of the Stream will refine our knowledge of its total gas mass. This could
be achieved by UV absorption studies of AGN lying far away on the sky
from the 21\,cm-emitting body of the Stream.

\item Further studies of the relationship between the \ha-emitting gas
and the UV-absorbing gas will help elucidate the properties
of the ionized gas, including the ionization mechanism.

\item Dynamical models based on cosmological simulations with gas dynamics
included, will greatly tighten
our constraints not only on the dynamics of the LMC-SMC but also on the
Local Group. It will be useful to further constrain how the Clouds are
accreted into the MW, their past orbital history and hence
scenarios for the formation of the Magellanic Stream.

\item The next big improvement in proper-motion measurements for the 
LMC and SMC will be possible with the \emph{Gaia} mission. 

\item Further \emph{HST} proper motion measurements 
will be extremely useful to better constrain 
the SMC's internal kinematics,
which are currently poorly understood. This will enhance our ability
to further constrain theoretical models of the SMC's dynamics.

\item In a few years, the proper motions of distant satellite galaxies in
our MW will be measured by the \emph{James Webb Space Telescope}.
This will likely confirm or confute the existence of a Magellanic
association of dwarf galaxies. 

\item Meanwhile, searches for Stream-analogs in high-redshift galaxies 
will help reveal whether the Magellanic Stream is unusual 
(in terms of gas mass and spatial extent).
We know from SDSS statistics that
the MW is somewhat unusual in having two relatively massive
dwarf companions so close by (Tollerud et al. 2011).
The nearby spirals M31 and M33 are connected by a bridge of H I 
(Braun \& Thilker 2004, Wolfe et al. 2013), although it is unclear
whether this represents a
condensing intergalactic filament rather than a tidal feature. More
examples of extragalactic gaseous tidal streams are needed to 
understand the role 
such features play in the larger picture of galaxy evolution.
\end{enumerate}
\end{issues}

We end this article by returning to the Stream's namesake, Ferdinand Magellan.
Although Magellan's crew completed the first circumnavigation of the globe
when they sailed his ship the {\it Victoria} back to Europe in 1522,
the voyage ended tragically for him in the Philippines.
There is a striking parallel between the around-the-world nature of his voyage
and the around-the-galaxy nature of the Magellanic Stream, revealed to us
almost 500 years later. 
Magellan's voyage opened up the world to trade and discovery.
His Stream is blazing its own trail around the Galaxy.

\section*{DISCLOSURE STATEMENT}
The authors are not aware of any affiliations, memberships, funding, or 
financial holdings that might be perceived as affecting the objectivity 
of this review.

\section*{ACKNOWLEDGMENTS}
E.D.O. acknowledges support from the Alfred P. Sloan foundation. 
We thank Jay Gallagher, Julio Navarro, Mary Putman, Snezana Stanimirovi\'c, 
Carlos Vera-Ciro, Bart Wakker, and Dennis Zaritsky for useful comments,
and Joss Bland-Hawthorn for an insightful review.
We thank Kat Barger, Gurtina Besla,  
Francois Hammer, Nitya Kallivayalil, and David Nidever for permission to 
use their graphics, Stephen Pardy for providing Figure 7,
and Vasily Belokurov for providing Figure 8.

\end{document}